\documentclass[preprint,11pt]{JHEP3} 

\JHEPspecialurl{http://jhep.sissa.it/JOURNAL/JHEP3.tar.gz}

\usepackage{epsfig,multicol,amsmath}

\def\beq{\begin{equation}}
\def\eeq{\end{equation}}
\def\bea{\begin{eqnarray}}
\def\eea{\end{eqnarray}}

\def\eq#1{{Eq.~(\ref{#1})}}
\def\fig#1{{Fig.~\ref{#1}}}
\newcommand{\bas}{\bar{\alpha}_S}
\newcommand{\as}{\alpha_S}

\newcommand{\Lb}{\left(}
\newcommand{\Rb}{\right)}

\parskip=\itemsep               

\newcommand{\h}{\frac{1}{2}}

\def\pom{{I\!\!P}}

\title{\LARGE \bf Soft processes at high energy  without soft Pomeron:\\
 a QCD motivated model.}
\author{\large  A. ~Kormilitzin\thanks{Email: andreyk1@post.tau.ac.il} \,\,\,and \,\,\, E. ~Levin\thanks{Email: leving@post.tau.ac.il,
levin@mail.desy.de;}  \\
Department of Particle Physics, School of Physics and Astronomy\\
Raymond and Beverly Sackler
 Faculty
of Exact Science\\  Tel Aviv University, Tel Aviv, 69978, Israel}

\abstract{In this paper we develop a QCD motivated model for  both
hard and soft interactions at high energies. In this model the long
distance behavior of the scattering amplitude  is determined  by the
approximate solution to the non-linear evolution equation  for
parton system in the saturation domain. All phenomenological
parameters for dipole-proton interaction were fitted from the deep
inelastic scattering data and the soft processes are described with
only one new parameter, related to the wave function of hadron. It
turns out that we do not need  to introduce the so called soft
Pomeron that has been used in high energy phenomenology for four
decades.  The model described all data on soft interactions: the
values of total, elastic and diffractive cross sections as well as
their $s$ and $t$ behavior. The value for the survival probability
of the diffractive Higgs production is calculated  being less 1\%
for the LHC energy range.  }


\preprint{  TAUP -2884-08\\
\today}
\begin{document}

\section{Introduction}
For three decades  our microscopic theory-QCD, has been unable to
describe soft (long distance) interactions. It happens, partly,
because of the embryonic stage of our  understanding of the most
challenging problem in QCD: confinement of quarks and gluons.  We
believe, however, that the main features of high energy scattering
at long distances can be described in the framework of QCD,  and such
a description is based on the phenomenon  of saturation of the
parton densities in QCD \cite{GLR,MUQI,MV}. Indeed, at short
distances the parton densities increase with the growth of energy
due to gluon emission, which is described by the QCD linear evolution
equations \cite{DGLAP,BFKL}.  Such an emission is proportional to
the density of emitters (partons).  The process of annihilation is
suppressed since it's strength is proportional to the parton density
squared, reflecting the fact that two partons have to meet each other
in the small volume of interaction. However,   at high energy the
density of partons becomes so large that the annihilation term tends
to be of the same order as the emission term. It leads to a slow down in
the increase of the parton density,  and the system of partons  lives
in the dynamical equilibrium with the critical value of density (it
is saturated).  We have several theoretical approaches for a QCD
description of the parton system in the saturation domain (see Ref.
\cite{GLR,MUQI,MV,BK,JIMWLK,KOLU,HIMST,STPH,EGM,LMP}, but the equations
that have been proposed turn out to be so complicated, that our
knowledge about this domain mostly stems from numerical solutions of
these equations \cite{NS}.

The alternative approach that has been developed during the past
decade  is  to build models \cite{MOD,KOR}  that incorporate the main
qualitative properties of  these solutions, such as the geometrical
scaling behavior of the scattering amplitude \cite{BALE,GS,IIM} and
the existence of the new dimensional scale (the so called saturation scale)
\cite{GLR,MUQI,MV}, which increases in the region of low $x$ (high
energy). Such models successfully described all the data on deep
inelastic scattering, including the behavior of inclusive and diffractive
cross sections, at a low value of photon virtuality \cite{MOD}.

The main goal of this paper is to describe the processes that occur
at long distances (soft processes), where we cannot apply
perturbative QCD, using the saturation model. In this model all the
needed phenomenological parameters, in  particular the  energy
behavior and the value of the saturation scale, were fitted using
the DIS data. The main ingredient of the high energy phenomenology
during the past four decades, namely the soft Pomeron, is not introduced in
this model. In simple words, we replace the soft Pomeron by the
scattering amplitude that describes the behavior of the saturated
parton system. The idea, that the matching between soft and hard
interaction occurs in the saturation region, and can be described by
the amplitude that originates from the high density QCD approach,
is not new. This idea has been in the air for a number of years  and
have been advocated in several lectures on the subject \cite{REV}.
The first attempt of a practical application of this idea was done
in Ref. \cite{BATAV} (see also Ref. \cite{SPLAST}). The result of
this paper was encouraging since the estimates gave a  reasonably
good agreement with the experimental data on the total cross section of the
pion, kaon, and proton interaction with the proton target.  In this
paper we continue  to develop these ideas and we give the comparison
with the experiment data  for the full set of the experimental data
of  soft interactions, that include the energy behavior of the total,
elastic and diffraction cross sections as well as a $t$- behavior of the
elastic cross sections, and the mass behavior for the cross sections
of  diffraction production.

\section{The main idea}
As has been mentioned, our main idea is to describe the so called
`soft' interaction in the high parton density QCD (hdQCD). In hdQCD
we are dealing with the dense system of partons which cannot be
treated in the perturbative QCD  approach. Being non-perturbative in its
nature, hdQCD leads to typical distances $r = 1/Q_s(x) \ll R \sim
1/\Lambda_{QCD}$ where $R$ is the size of a hadron. The saturation
momentum $Q_s(x)$ is a  new dimensionful scale  which increases
with the growth of the energy ( or at $x \to 0$)\cite{GLR,MUQI,MV}.
Therefore, the QCD coupling $\as\Lb Q_s(x)\Rb$ is small and, because
of this, we are able to develop a theoretical approach for such a
dense system having only a limited input from the unknown
confinement region.  To illustrate the limitation of our approach, it
is enough to look at the expression for the saturation scale in
hdQCD\cite{GLR,QS} \bea \label{TEQS}
Q^2_s(Y)\,\,&=&\,\,Q^2_s(Y_0)\,\exp  \left( \bas
\,\frac{\chi(\gamma_{cr})}{1 - \gamma_{cr}}\,\,(Y - Y_0)\, \,-
\,\,\frac{3}{2 ( 1 - \gamma_{cr})}\,\ln(Y/Y_0)
-  \right.  \nonumber\\
\, &-&\,\left. \frac{3}{( 1 -
\gamma_{cr})^2}\,\sqrt{\frac{2\,\pi}{\bas\,\chi''(\gamma_{cr})}}\, (
\frac{1}{\sqrt{Y}}\,-\, \frac{1}{\sqrt{Y_0}}\,)\, ,+\,O(\frac{1}{Y})
\right) \eea In \eq{TEQS}  the energy dependence of
$Q_s(Y=\ln(1/x))$  is predicted from hdQCD, ($\chi(\gamma)$ is the
Mellin transform of the BFKL kernel) but the scale $,Q^2_s(Y_0)$
stems from the confinement region, demonstrating  a need for some
input from this theoretically unknown region. Our hope that such
input will be a limited number of constants.

Let us consider the deep inelastic electron-proton  scattering (DIS),
to illustrate our point.  If the virtuality of the photon $Q^2$ is very
large, $Q^2 \gg  Q^2_s(x)$, the typical distances $r \approx 1/Q $, and
we can safely  apply  the perturbative QCD  approach based on the
operator product expansion, and the DGLAP evolution equation. However, if
$ \Lambda^2_{QCD} \ll Q^2 \ll Q^2_s(x)$, the situation changes
crucially. All terms in the operator product expansion become of the
same order, and we have to use the hdQCD approach.  In this approach,
the cross section $\sigma^{\gamma^* p}$ shows a geometrical scaling
behaviour\cite{GESC}, namely, $\sigma^{\gamma^* p} = \Lb
1/Q^2_s(x)\Rb\,F\Lb Q^2/Q^2_s(x)\Rb$ which has been confirmed
experimentally.  This means that the typical distances $ r \approx
1/Q_s(x) \ll 1/\Lambda_{QCD}$, which are short.

Our idea is that we can describe  $ Q^2  \approx \Lambda^2_{QCD}$,  in
the framework of the same hdQCD approach.  Having the geometrical
scaling behavior in mind, such an idea does not look crazy. However,
in DIS we are dealing with the total cross sections which are
related to the amplitude integrated over the impact parameters ($b$).
For a treatment of  the soft interaction observables, we need to know the
$b$ dependence. As was shown in Ref. \cite{KOWI}, hdQCD  predicts the
power-like decrease at large value of $b$, which contradicts both the
theoretical estimates and experimental observations. Therefore, we
need an input from the confinement region to specify the $b$
dependence of the scattering amplitude.  Let us first discuss large
$b \geq 1/\Lambda_{QCD} \gg 1/Q_s(x)$. The scattering amplitude for
DIS $A\Lb Q^2,W;t\Rb$  depends on the photon virtuality $Q^2$, energy $W
= \sqrt{s}= Q/\sqrt{x}$ and on the momentum transferred $t = -q^2 <0$.
Using the dispersion relation in the $t$-channel we have \beq \label{DR}
A\Lb
Q^2,W;t\Rb\,\,=\,\,\frac{1}{\pi}\,\int^{\infty}_{4m^2_\pi}\,\frac{d
t' \,Im_t A\Lb  Q^2,W;t'\Rb}{t' + q^2} \eeq Calculating the
amplitude in the impact parameter representation we can reduce
\eq{DR} to the form \bea \label{DR1}
A\Lb Q^2,W;b\Rb\,\,&=&\,\,\frac{1}{(2 \pi)^2} \,\int\,d^2\vec{q} \,A\Lb Q^2,W;t\Rb\,e^{i \vec{q}\cdot \vec{b}}\,\,=\,\,\frac{1}{2\pi^2}\,\int^{\infty}_{4m^2_\pi}\,d t'\, Im_t A\Lb  Q^2,W;t'\Rb\,\int\,\frac{q\,d q\,\,J_0\Lb q b \Rb}{t' + q^2}\,\\
&=&\,\,\frac{1}{2\pi^2}\,\int^{\infty}_{4m^2_\pi}\,d t' \,K_0\Lb \sqrt{t'}\,b\Rb\,\,Im_t A\Lb  Q^2,W;t'\Rb\,\,\
\stackrel{b \geq 1/(2 m_\pi)}{\longrightarrow}\,C\,Im_t A\Lb  Q^2,W;t'=4m^2_\pi\Rb\,K_0 \Lb 2 m_\pi b \Rb\nonumber
\eea
where in the constant $C$, all the numerical factors  have been absorbed.

In \eq{DR1}, the $b$-dependence is determined by the mass of the
lightest hadron (pion), and cannot not be reproduced in perturbative
or/and high density QCD. However, one can see from \eq{DR1} that we
have a factorization of the $b$ dependence, and the dependence on the photon
virtuality and energy.

As it is well known (see Ref. \cite{LMP} and references therein), in
the kinematic region $\as \,\leq \as\,\ln(1/x) \leq 1/\as$ in the
saturation region the hdQCD approach reduces to the BFKL Pomeron
calculus. We generalize \eq{DR1} for the BFKL Pomeron in the
following way \beq \label{DR2} A_{BFKL}\Lb Q^2,x;b\Rb\,\,\,=\,\,\,
\stackrel{\mbox{ \small short distances}}{A_{BFKL}\Lb
Q^2,x;t=0\Rb}\,\times\,\stackrel{\mbox{\small long distances}}{S(b)}
\eeq One can see that \eq{DR2} claims a kind of factorization
between short and long distances. The short distance part we can
describe in hdQCD, while the long distance part has to be taken from
the confinement domain. In a more general way, we can re-write \eq{DR2} in
the form \beq \label{DR3} A_{BFKL}\Lb Q^2,x;\vec{b}\Rb\,\,\,=\,\,\,
\int d^2\vec{b'} \stackrel{\mbox{ \small short
distances}}{A^{}_{BFKL}\Lb Q^2,x;
\vec{b'}\Rb}\,\times\,\stackrel{\mbox{\small  long
distances}}{S(\vec{b} - \vec{b}' )} \eeq where $b'$ in the short
distance part is of the order of $b' \approx 1/Q_s(x)$.

The approach of \eq{DR2} and \eq{DR3}, at first sight contradicts
the high energy phenomenology based on the soft Pomeron and Reggeons, as
well as well as the lattice calculations \cite{LAT}. In both
approaches, the Pomeron trajectory has $\alpha'_{\pom}  > 0$, and at
positive $t$ the glueballs lie on this trajectory. In \eq{DR2} and
\eq{DR3}, the Pomeron slope $\alpha'_{\pom}  =0$.  We will show below
that we will be able reproduce the experimental data on energy
dependence of the elastic slope which used to consider as the
argument for $ \alpha'_{\pom} = 0.25 \,GeV^{-2}$.  On the theoretical side,
the only theory which can treat at the moment on the same footing
both short and long distances: N=4 SYM with AdS/CFT
correspondence, leads to a picture which is in striking agreement with
our approach\cite{BST}. Namely, in this approach at $t>0$ (
resonance region)  we have a normal soft Pomeron with
$\alpha'_{\pom}  > 0$ while at $t <0$ (scattering region)
$\alpha'_{\pom}  = 0$.

\section{The model}

\subsection{Motivation}

Our building of the model is based on the dipole approach to high
energy QCD \cite{MUCD}.  In this approach, the evolution of the
parton (colorless dipole) system  can be written in the form of the
equation for the generating functional, with  a transparent
probabilistic interpretation for them. The generating functional is
defined as \cite{MUCD} \beq \label{Z} Z\left(Y\,-\,Y_0;\,[u]
\right)\,\, \equiv\,\,\sum_{n=1}\,\int\,\, P_n\left(Y\,-\,Y_0;\,x_1,
y_1; \dots ; x_i, y_i; \dots ;x_n, y_n
 \right) \,\,
\prod^{n}_{i=1}\,u(x_i, y_i) \,d^2\,x_i\,d^2\,y_i \eeq where $u(x_i,
y_i) \equiv u_i $ is an arbitrary function of $x_i$ and $y_i$. The
coordinates $(x_i,y_i)$ describe the colorless pair of gluons or a
dipole. $P_n$ is a probability density to find $n$ dipoles with the
size $x_i - y_i$, and with impact parameter $(x_i + y_i)/2$.
Assuming that we have only a decay of one dipole to two dipoles,
directly from the physical meaning of $P_n$  it follows
\cite{MUCD,L1,L2} \beq\label{P} \frac{\partial
\,P_n(Y;\dots;x_i,y_i; \dots;x_n,y_n)}{\partial Y}\,\,\,=
 \eeq
$$
 =\,\, \sum_{i} \  V_{1 \to 2}
\bigotimes \left(P_{n-1}(Y; \dots;x_i,y_i; \dots;x_n,y_n)\, -\,P_n (Y;\dots;x_i,y_i;
\dots;x_n,y_n)\right)
$$
where $\bigotimes$ denotes all necessary integrations and $V_{1 \to
2}$ is the vertex for the decay of one dipole to two dipoles. This
vertex is equal to \beq \label{V} V_{1 \to 2}\Lb (x,y) \to (x,z) +
(z,y) \Rb \,\,=\,\,\frac{\bas}{2 \pi}\,\frac{( \vec{x} -
\vec{y})^2}{(\vec{x} - \vec{z})^2\,(\vec{z} - \vec{y})^2} \eeq
\eq{P} is a typical Markov's chain which takes into account the $s$
- channel unitarity on each step of evolution since it has two
terms: the  birth of the new dipole due to the decay of the parent
dipole (positive term in \eq{P}) and the death of the dipole due to
the same process.

\eq{P} can be rewritten as the equation for the generating
functional \cite{L1,L2}, namely, \beq \label{ZEQ} \frac{\partial
\,Z\,\Lb Y-Y_0; [\,u\,]\Rb}{\partial \,Y}\,\,= \,\,\int\,d^2 x\,d^2
y\,d^2 z\,\, V_{1 \to 2}\Lb (x,y) \to (x,z) +
(z,y)\Rb\,\frac{\partial}{\partial \,u(x,y)}\,\,Z\,\Lb Y- Y_0;
[\,u\,] \Rb
  \eeq
$Z\,\Lb Y-Y_0; [\,u\,]\Rb$ satisfies the initial and boundary conditions:
\bea
\mbox{Initial\,\,\,\,\,\,\,\,\,conditions:} & Z\,\Lb Y-Y_0 = 0; [\,u\,]\Rb & =\,\,\,u(x,y)\,; \label{IC} \\
\mbox{Boundary conditions:} & Z\,\Lb Y-Y_0 ; [\,u = 1\,]\Rb &
=\,\,\,1\,; \label{BC} \eea The advantage of the generating
functional is that we can write it in terms of this simple functional
formulae for the scattering amplitude, which has an obvious partonic
interpretation. Indeed, the scattering amplitude in the lab. frame
can be written in the form \cite{BK,L1,L2} \bea \label{N}
&& N\Lb Y - Y_0; r,b \Rb \,\,=\\
&&\,\,\sum^{\infty}_{n =1}\,\frac{(-1)^n}{n!}\,\,\int
\,\prod^{n}_{i=1}\, \,d^2\,x_i\,d^2\,y_i\,\,\rho_n \Lb
r,b;\{x_i,y_i\}, Y - Y_0\Rb \gamma_n \Lb r,b;\{x_i,y_i\}, Y_0 \Rb
\nonumber \eea where  $Y - Y_0\,\,\gg\,\,1$ but
$Y_0\,\,\approx\,\,1$  and \beq \label{RHO} \rho_n \Lb Y -
Y_0,\{x_i,y_i\} \Rb\,\,\,=\,\,\,\prod^n_{i =1}\,\frac{\delta}{\delta
u(x_i,y_i)} Z\,\Lb Y-Y_0; \{\,u(x_i,y_i)\,\}\Rb|_{u(x_i,y_i)=1} \eeq
and $\gamma_n$ is the  amplitude of interaction of $n$ dipoles at
low energy ( small values of rapidity $Y_0$) with the coordinates
$x_i,y_i$ with the target with size $r$ and impact parameter $b$. In the
parton language the generating functional determines the parton wave
function for which we have the evolution equation \eq{ZEQ} in  QCD.
$\gamma_n$ can have a non-perturbative origin. We assume that
$\gamma_n \Lb r,b;\{x_i,y_i\}, Y_0 \Rb\,\,=\,\,\prod^{n}_{i=1}
\gamma(x_i,y_i;Y_0)$.  This assumption means that the dipoles inside the
target have no correlation, which is correct, as far as we know, only
for a nucleus target.

In spite of the transparent physics behind our equations, the equation
for the generating functional is difficult to solve  analytically,
especially if we generalize \eq{ZEQ} to the case of the so called
Pomeron loops \cite{MSHW,LELU,IT}.  As has been mentioned, we wish
to find a model solution to \eq{ZEQ}. We will do this using  two key
observations of Ref.\cite{LMP}. The first one is the fact that in the
huge kinematic region of $\bas Y\,\leq 1/\bas$,  the system of
partons that we are dealing with  turns out to be a system of
non-interactive partons. In other words, it means that the generating
functional of \eq{Z} can be rewritten in a simpler form, namely.
\beq \label{ZEQF}
 Z\,\Lb Y-Y_0; \{\,u(x_i,y_i)\,\}\Rb\,\,\,=\,\,\,\sum^{\infty}_{n=1}\,\frac{C_n}{n!}\,\,\Lb \int \,d^2 x'\,d^2 y'\,P(Y-Y_0;x,y;x',y')\,\,\Lb u(x',y')\,-\,1\Rb\Rb^n
\eeq where $P(Y-Y_0;x,y;x',y')$ is the amplitude for one BFKL
Pomeron  exchange, between dipoles $(x,y)$ and $(x',y')$.

In Ref.\cite{LMP} it was shown  , using the analytical solution of Ref. \cite{LT} , that
actually $C_n\,=\,1$ for the dipoles with the size $r \sim 1/Q^2_s$ where $Q_s$ is the saturation momentum.

Therefore, \eq{ZEQF} can be simplified and rewritten in the form
\beq \label{EIKZ}
 Z\,\Lb Y-Y_0; \{\,u(x_i,y_i)\,\}\Rb\,\,\,=\,\,\,\exp\Lb  \int \,d^2 x'\,d^2 y'\,P(Y-Y_0;x,y;x',y')\,\Lb  u(x',y')\,-\,1
\Rb\Rb
\eeq

and the amplitude has the form
\beq \label{EIKN}
N\Lb Y-Y_0; x,y\Rb\,\,=\,\,1 - \exp \Lb - \int\,d^2 x' \,d^2 y'\,\, P\Lb Y - Y_0;x,y; x',y'\Rb \gamma \Lb Y_0; x',y'\Rb \Rb
\eeq
where $\gamma \Lb Y_0; x',y'\Rb$  is the scattering amplitude of the dipole $( x',y')$, with the target at low energy.

\subsection{Main formulae and assumptions}

\eq{EIKN} is the main  formula for constructing our model.  The first
ingredient of the model is the choice of the amplitude $\gamma$ and
the value of $Y_0$. We choose this amplitude  in the form \cite{QXS}
\beq \label{M1} \gamma \Lb Y_0; x',y'\Rb \,\,=\,\,\gamma_{mod} \Lb
Y_0; x',y'\Rb \,\,=\,\, \,\,\frac{\pi^2\,\as(\mu^2)}{3}\,r^2\,x_0
G^{DGLAP}(x_0,\mu^2)\,S(b) \eeq where $ \vec{r}\,=\,\vec{x} -
\vec{y}$ and $\vec{b} = \h\,(\vec{x} +  \vec{y})$ with $x_0 =
10^{-2}$. $xG$ in \eq{M1} is the solution of the DGLAP evolution
equation which describes the DIS experimental data. $\mu$ is equal
to \beq \label{MU} \mu^2\,\,=\,\,\mu^2_0\,\,+\,\,\frac{C}{r^2} \eeq
where $\mu_0$ and $C$ are phenomenological parameters which has to
be found fitting the experimental data on DIS (see Ref.\cite{KOR}
for details). \eq{MU} means that $\gamma_{mod} \Lb Y_0; x',y'\Rb$ at
large dipole size $ r \mu_0 \gg 1  $ is determined by the following
expression \beq \label{LRD} \gamma_{mod} \Lb Y_0; x',y'\Rb \,\,=\,\,
\,\,\frac{\pi^2\,\as(\mu^2_0)}{3}\,r^2\,x_0
G^{DGLAP}(x_0,\mu^2_0)\,S(b) \eeq Of course \eq{LRD}  has no
theoretical basis, and our hope is that the amplitude at low $x$ will be
not very sensitive to this kinematic region, in the initial
condition.

$S(b)$ is the impact parameter profile function, which is chosen as the Fourier
image of the proton form factor, namely,
\beq \label{SB}
S(b)\,\,=\,\,\frac{2}{\pi \,R^2}\Lb \frac{\sqrt{8}\,b}{R} \Rb\,K_1 \Lb \frac{\sqrt{8}\,b}{R} \Rb
\eeq
where $R$ is the proton radius ($R = 0.89\,fm$).

In \eq{EIKN} $ P\Lb Y - Y_0;x,y; x',y'\Rb$ stands for the BFKL
Pomeron Green's function. This Green's function describes the energy
($x$)  evolution of the gluon system in the region of  high energy
(low $x$). This evolution allows us to find the dipole scattering
amplitude or parton density at low $x$ from the initial condition:
the amplitude at lower $x = x_0$, but for any value of the dipole size
($r$).

However, we know that the BFKL Pomeron alone cannot describe the
experimental data, since it determines the anomalous dimension only
at  low $x$. The experimental data shows a good agreement with the
DGLAP evolution equation, which is written as the evolution in
$\ln(\mu^2)$ (see \eq{M1}), which gives us the parton density from
the boundary condition: the parton density at $\mu^2 = \mu^2_0$ but
at any value of $x$ including the region of low $x$. Strictly
speaking, it  cannot be used in \eq{EIKN}, but assuming \eq{MU} for
the scale $\mu^2$, we impose the condition that at long distances, the typical
momentum scale in the linear evolution equation is equal to
$\mu^2_0$. In the framework of this assumption, it looks reasonable
to assume that \bea \label{OMEGA}
\Omega(Y,r,b)\,\,\,&=&\,\,2 \,\int\,d^2 x' \,d^2 y'\,\, P\Lb Y - Y_0;x,y; x',y'\Rb \gamma \Lb Y_0; x',y'\Rb \,\,\nonumber \\
 &=& \frac{\pi^2\,\as(\mu^2)}{3}\,r^2\, xG^{DGLAP}(x,\mu)\,S(b)
\eea

Finally, we can write the following model expression for the scattering amplitude;
\beq \label{MODN}
N\Lb x; r,b \Rb\,\,\,=\,\,1 \,\,\,-\,\,\exp\Lb - \h \Omega\Lb x,r,b \Rb\Rb
\eeq
where $\Omega\Lb x,r,b \Rb$ is given by \eq{OMEGA}.

\subsection{Fixing the values of the phenomenological parameters}

In \eq{MU} we have introduced two phenomenological parameters:
$\mu^2_0$ and $C$. The first determines the value of  the virtuality
of the probe in DIS processes from which we can start  using the
perturbative QCD evolution equations. The second ($C$) gives the
relation between scales in the momentum representation and in the coordinate
representation. In the case of the DGLAP evolution, $C = 4$ ( see Ref.
\cite{QXS}), but since the amplitude has a more complicated form than
\eq{MODN}, the value of  $C$ can be different.

Two more parameters stem from the initial condition for the gluon
density in \eq{OMEGA}, namely, \beq \label{IC1} xG^{DGLAP}\Lb x,
Q_0^2\Rb\,\,\,=\,\,\frac{A}{x^{\omega_0}}(1 - x)^6\,\eeq which
corresponds to the exchange of the soft Pomeron with the intercept
$\omega_0$ at the initial hard scale $\mu^2 = Q^2_0$ in traditional high
energy phenomenology, and the factor $(1-x)^6$ reflects the valence
quark dependance. In our approach we do not assume that the soft Pomeron
exists, and we view \eq{IC1} just as the simplest function that
reflects the behavior of the experimental data in DIS.

We use the observation of Ref. \cite{EKL}, that the anomalous
dimension in the leading order can be written in a very simple form,
namely, \beq \label{FP1} \gamma( \omega)\,\,=\,\,\bas \Lb
\frac{1}{\omega}\,\,-\,\,1 \Rb \eeq The explicit solution to the
DGLAP evolution equation with $\gamma(\omega)$ of \eq{FP1}, and with
the initial condition of \eq{IC1}, has been found in Ref. \cite{KOR}
and looks as follows \beq \label{FP2} xG^{DGLAP}(x,
t)\,=\,\sum_{k=0}^{6}
\binom{6}{k}\,(-1)^{k}\,xG^{(k)}(x,t,\omega_k)\eeq where \beq
\label{FP2} xG^{(k)}\Lb y\equiv \ln(1/x), t, \omega_{k}\Rb
\,\,=\,\,A\,e^{ - t \,+\,\omega_k\,y}\,\,\left\{\int^y_0\,\,d y'
\,e^{-\omega_k\,y'}\,\sqrt{\frac{t}{y'}}\,\,I_1 \Lb    2
\sqrt{t\,y'} \Rb \,\,\,+\,\,\,1 \right\} \eeq

with \beq \omega_k = \omega_0 - k\,; \;\;\;\; \label{t}
t\,\,\,\equiv\,\,\,\frac{4 N_c}{b_0}\,\ln \frac{\ln \Lb
\mu^2/\Lambda^2\Rb }{\ln \Lb
\mu^2_0/\Lambda^2\Rb}\,;\,\,\,\,\,\,\,\, b_0\,\,=\,\,11 - \frac{2
n_f}{3}\,\,\, \,\,\mbox{and}\,\,\,\,\,\,\,\,y\,=\,\ln(1/x) \eeq In
\eq{OMEGA} the typical scale of hardness in the gluon structure
function is determined by the process of gluon emission while the
factor $\as\,r^2$ takes into account the integration over the wave
function of the dipole with the size $r$.  Having this in mind, we
introduce a different scale for $\as$ in \eq{OMEGA}. Finally, we use
\beq \label{OMEGAPH} \Omega(x,r,b)\,\,\,
 =\,\,\, \frac{\pi^2\,\as(\tilde{\mu}^2)}{3}\,r^2\, xG^{DGLAP}(x,\mu)\,S(b) \,\,\,\mbox{with}\,\,\,
\tilde{\mu}^2\,\,=\,\,\tilde{\mu}^2_0\,\,+\,\,\frac{C}{r^2} \eeq
where $xG^{DGLAP}(x,\mu)$ is given by \eq{FP2} and the initial hard
scale $\widetilde{\mu}^{2}_{0}$ is different from that of
$\mu_{0}^{2}$. We perform a fit of the saturation model, to the DIS
experimental data and fixing the parameters, we wish  to describe
the soft data without using the concept of the "soft" pomeron. The fit
procedure was performed by using the minuit routine, the statistical and
systematical errors have been added in quadrature. We have used all of the
recent data on deep inelastic scattering processes, from different
collaborations. The most suitable were H1 \cite{Adloff:2000qk} and
ZEUS \cite{Chekanov:2001qu,Breitweg:2000yn}. It was observed, that
H1 data for medium and large values of $Q^{2}$ coincides with that
from the ZEUS col., but up to a scaling factor of 1.05. Hence, for the
fitting procedure, we took only the ZEUS data for vary small, medium and
large values of $Q^{2}$. Since we are interested in the high energy
description, we took data below $x<0.01$ and $0.045<Q^{2}<150
GeV^{2}$. The upper limit on virtuality is originated from the upper
limit on the $x$ variable. Fitting with H1 scaled data, yields almost
the same result. The total number of experimental points was 170.
During the fitting procedure, we observed a strong correlation
between different parameters, so we decided to fix a number of them.
We choose to fix a parameter $C$ and the initial hard scale
$Q^{2}_{0}$. The masses of the quarks were taken as follows:
$m_{light}\;=\;0.25\;GeV$ and $m_{charm}\;=\;1.3\;GeV$. $\chi^2$ for
this parametrization is $\thicksim 1.05$. The summary of the fitting
procedure is given in the table \ref{sum_tab}

\TABLE[ht]{
\begin{tabular}{|c|c|c|c|c|c|c|c|} \hline
Model & $\widetilde{\mu}_{0}^2$    &   $\mu_{0}^2$  & $Q_{0}^{2}$  & $C$ & $\omega_{0}$  &  $A$  & $\chi^{2}/d.o.f.$ \\
\hline \hline
A & 0.23 & 1.23 & \textbf{1.0} & \textbf{1.0} & 0.028 &
1.81 & 1.04\\ \hline \hline
B &0.069  & 1.23 & \textbf{0.58} & \textbf{1.0} & 0.087 &0.6
 & 3.08  \\ \hline
\end{tabular}
\caption{\label{sum_tab}\emph{Resulting parameters from the fit to
DIS data Model A  gives a very good description of the DIS data, while
model B reproduces quite well the data on soft processes}.} }

The bolded font corresponds to the fixed parameters $C$ and
$Q^{2}_{0}$, which were chooses to be equal to $1.0$. The value of the
hard scale $\mu^{2}_{0},\; \widetilde{\mu}_{0}^{2}$. is in the units of
energy $GeV$.

As one can see, $\chi^2/d.o.f.$ is close to 1 and fit gives very good
description of all experimental data on DIS (see \fig{dis}).
However, we present in Table 1 a second fit which leads to worse
$\chi^2/d.o.f.$ but reproduces the data on soft interaction quite
well as we will see below. It should be stressed that both fits
describe the DIS data at large values of photon virtualities with
small $\chi^2/d.o.f.$ . Therefore, we can consider model the A  as an
attempt to describe the  experimental data on DIS and soft processes,
assuming that \eq{MODN} is correct in the saturation region.  In
model B we explore a different idea: \eq{MODN}  is only approximate
formula that incorporates the main qualitative properties of
scattering amplitude for both DIS and soft scattering in the
saturation region. The question which we try to answer using  model
B is the following: is it possible to give the unique description of
long distance physics both in DIS and soft interaction based on the
QCD amplitudes at short distances.

\FIGURE[h]{\begin{tabular}{c c c}
\epsfig{file= 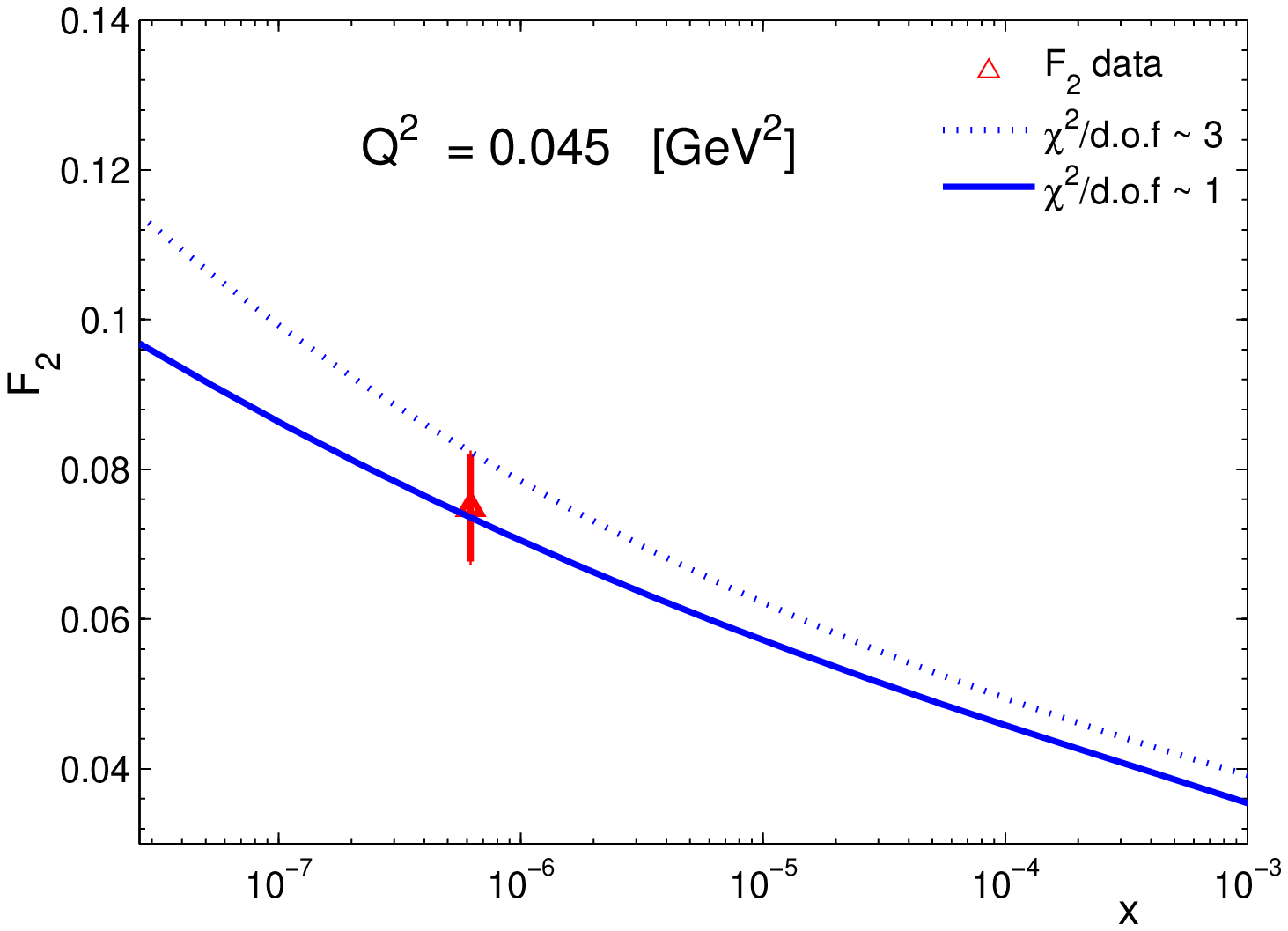,width=55mm} &\epsfig{file= 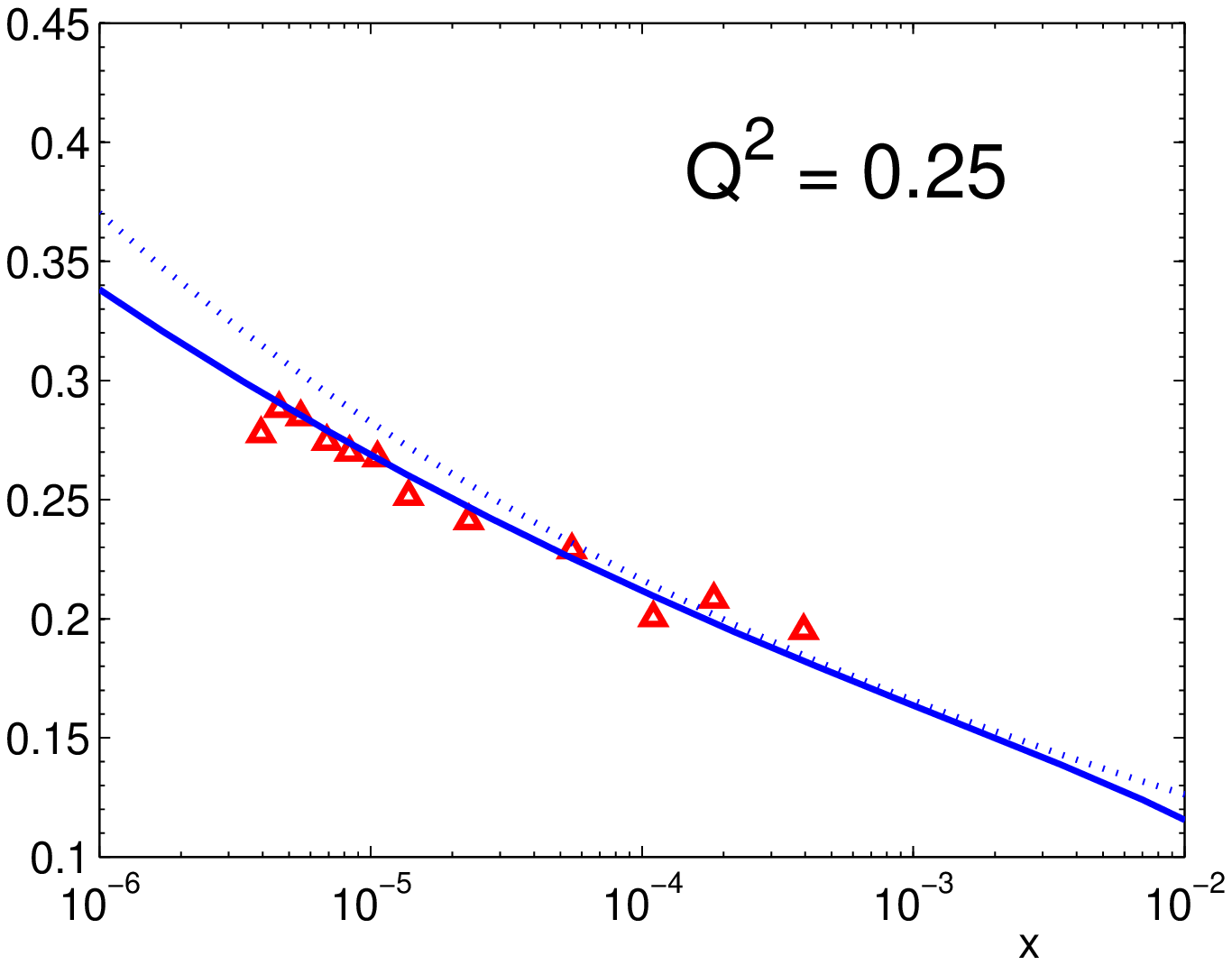,width=55mm} &\epsfig{file= 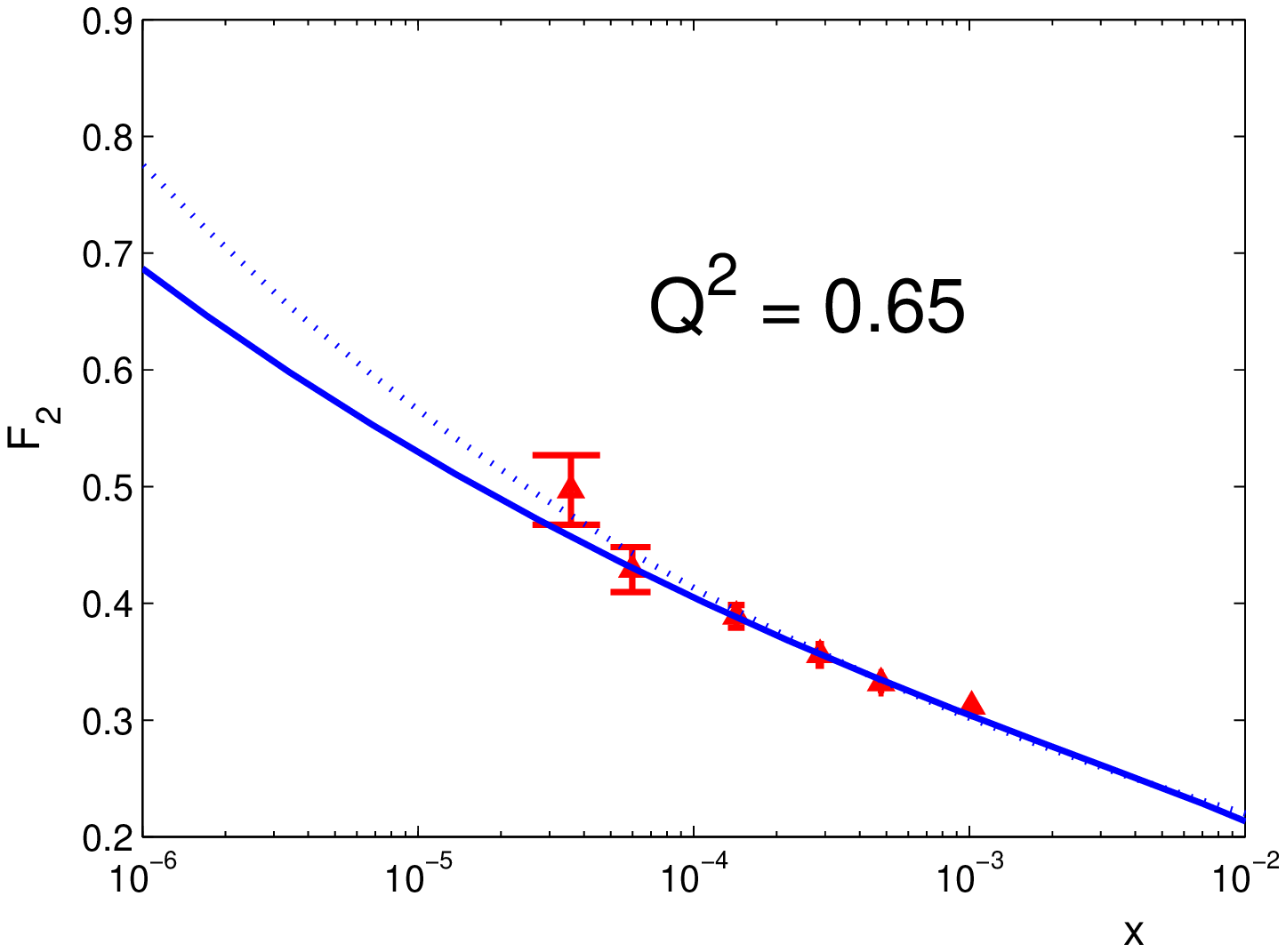,width=48mm, height=40mm}\\
\epsfig{file= 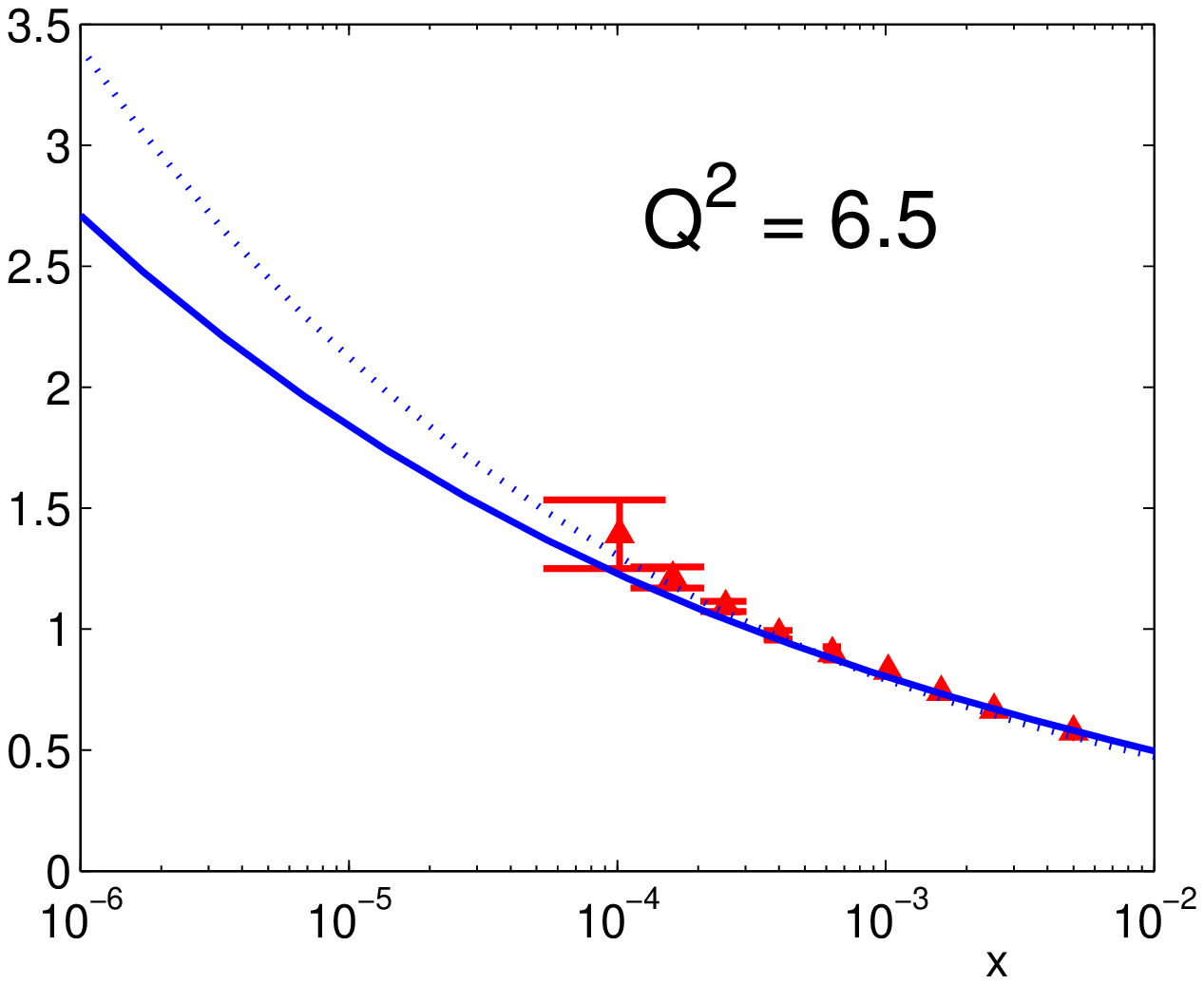,width=55mm} &\epsfig{file= 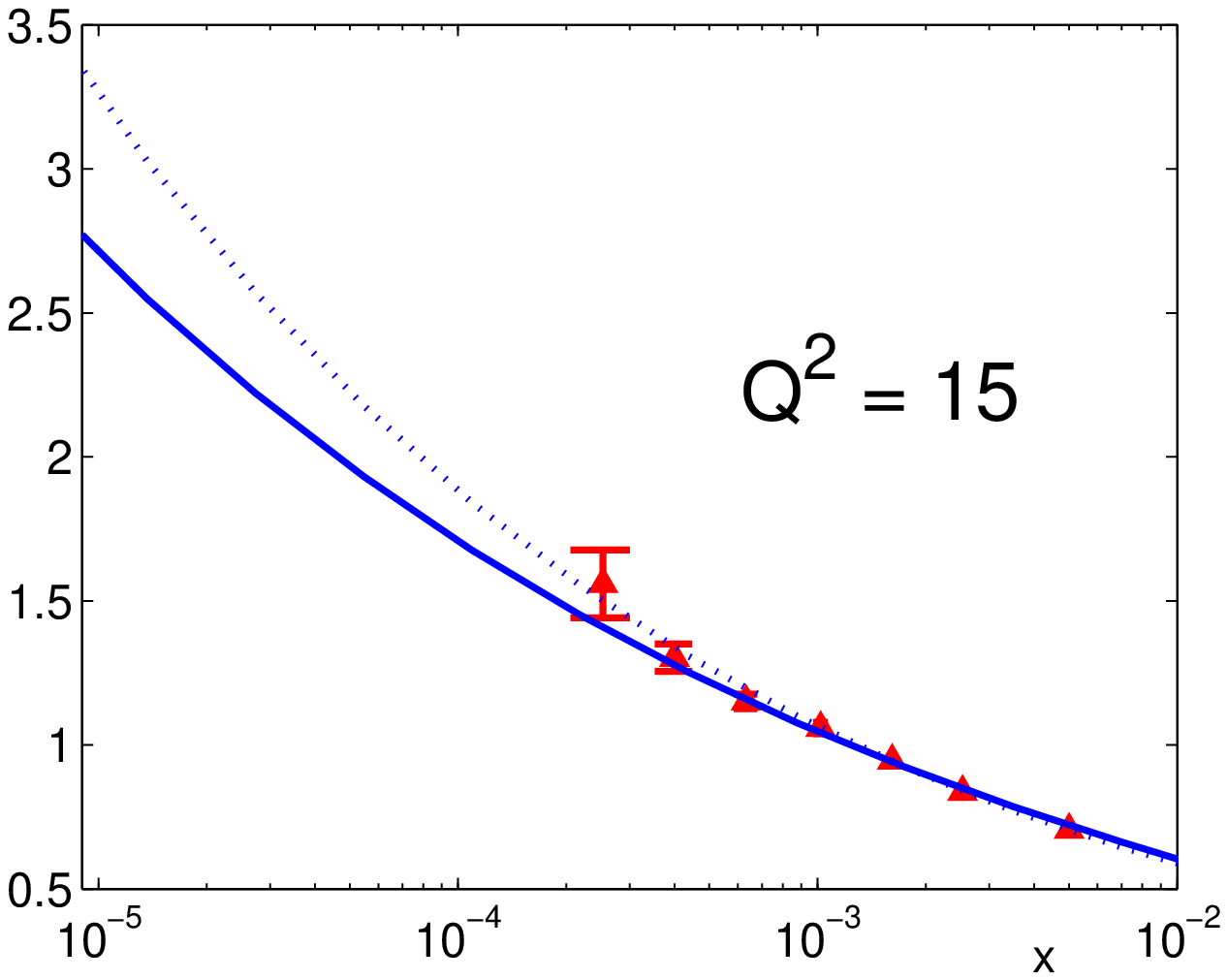,width=55mm} &\epsfig{file= 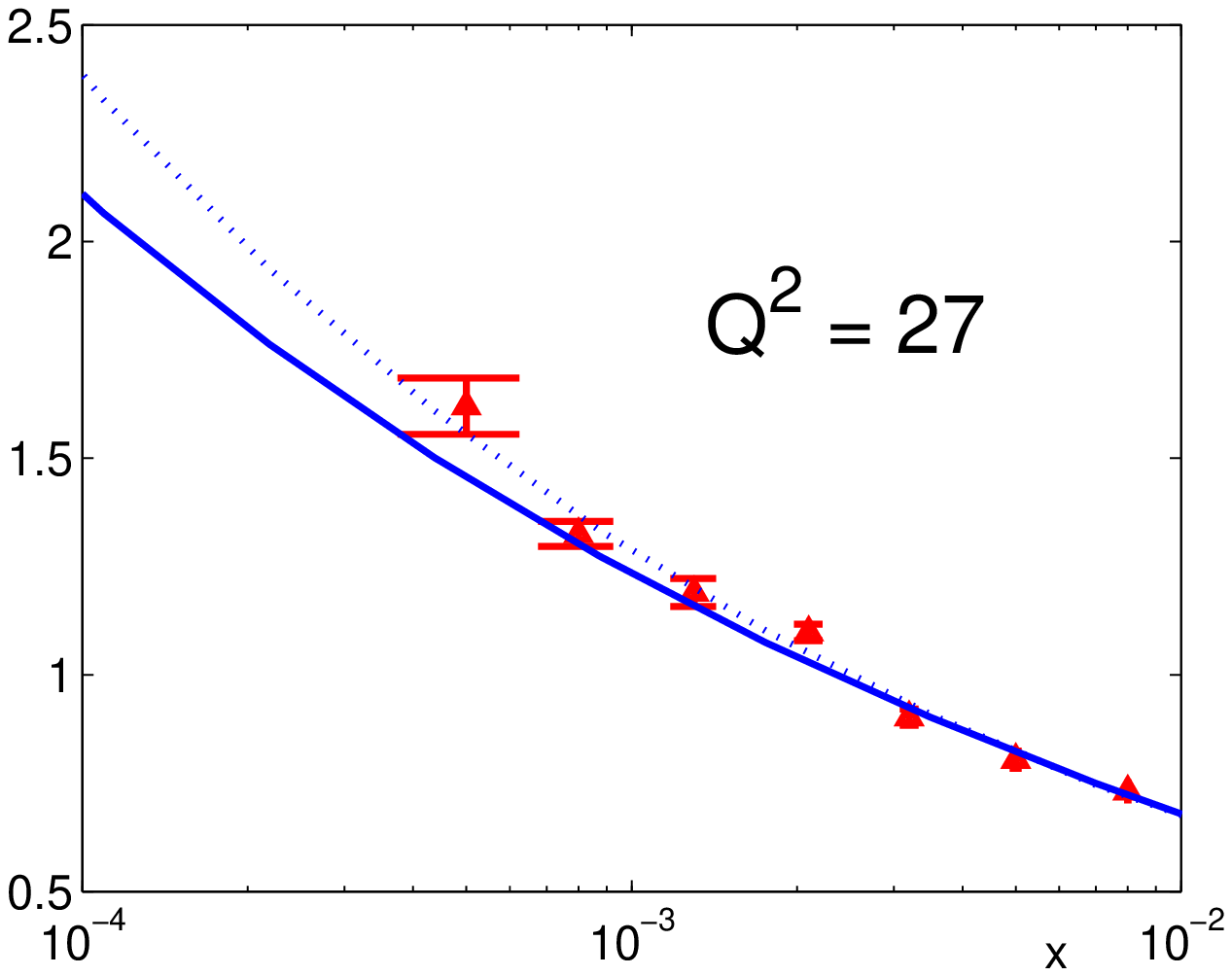,width=55mm}\\
\epsfig{file= 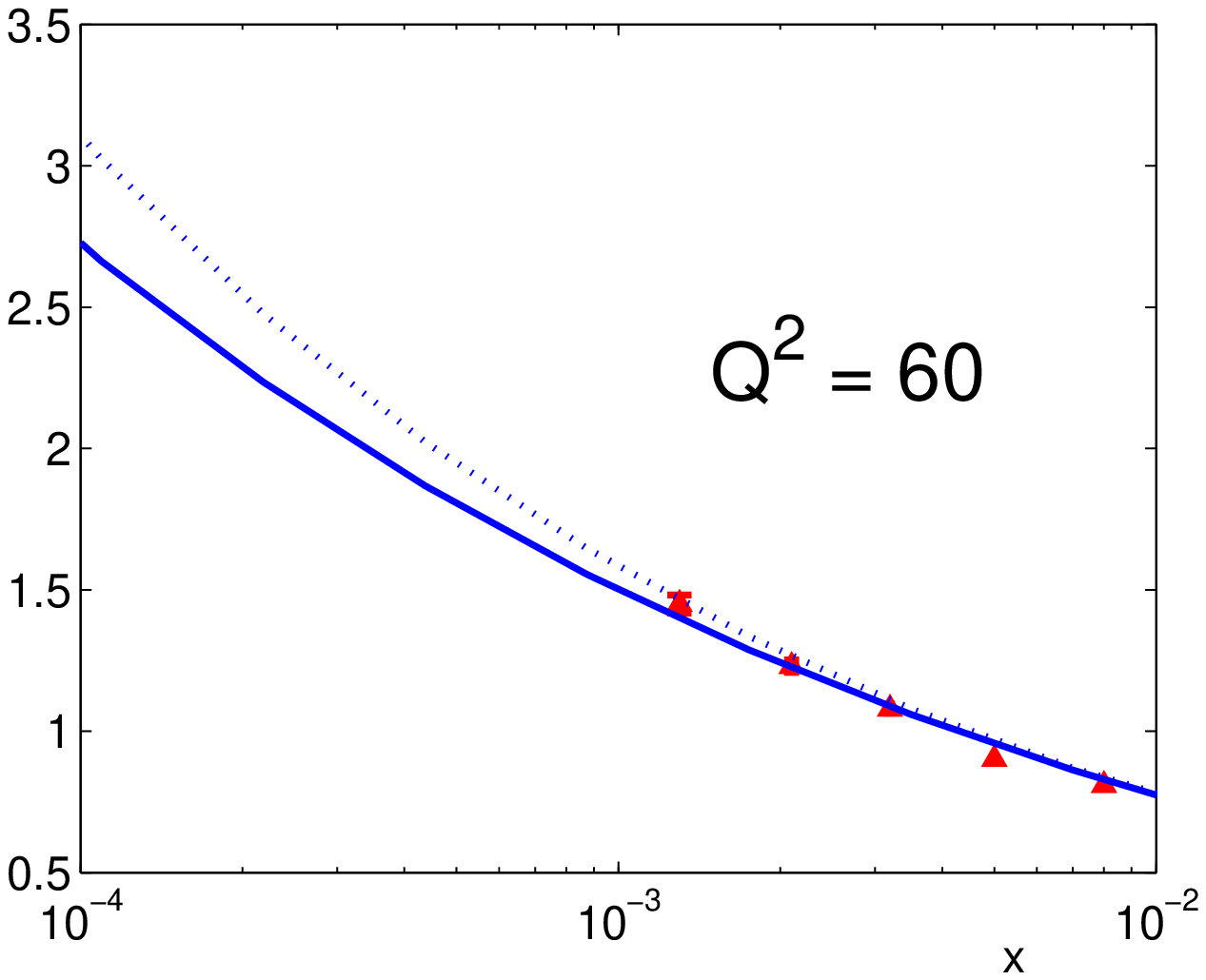,width=55mm} &\epsfig{file= 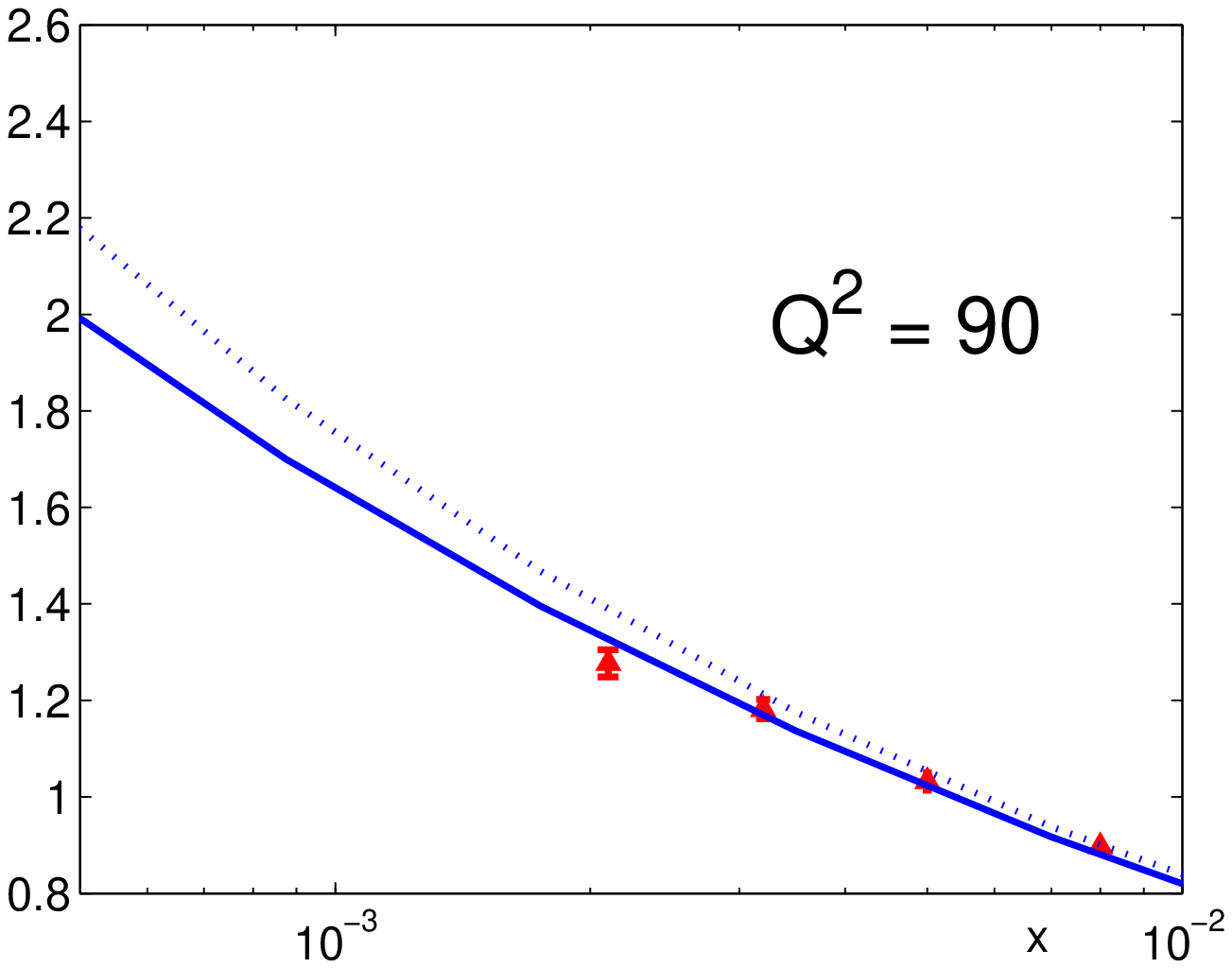,width=55mm} &\epsfig{file= 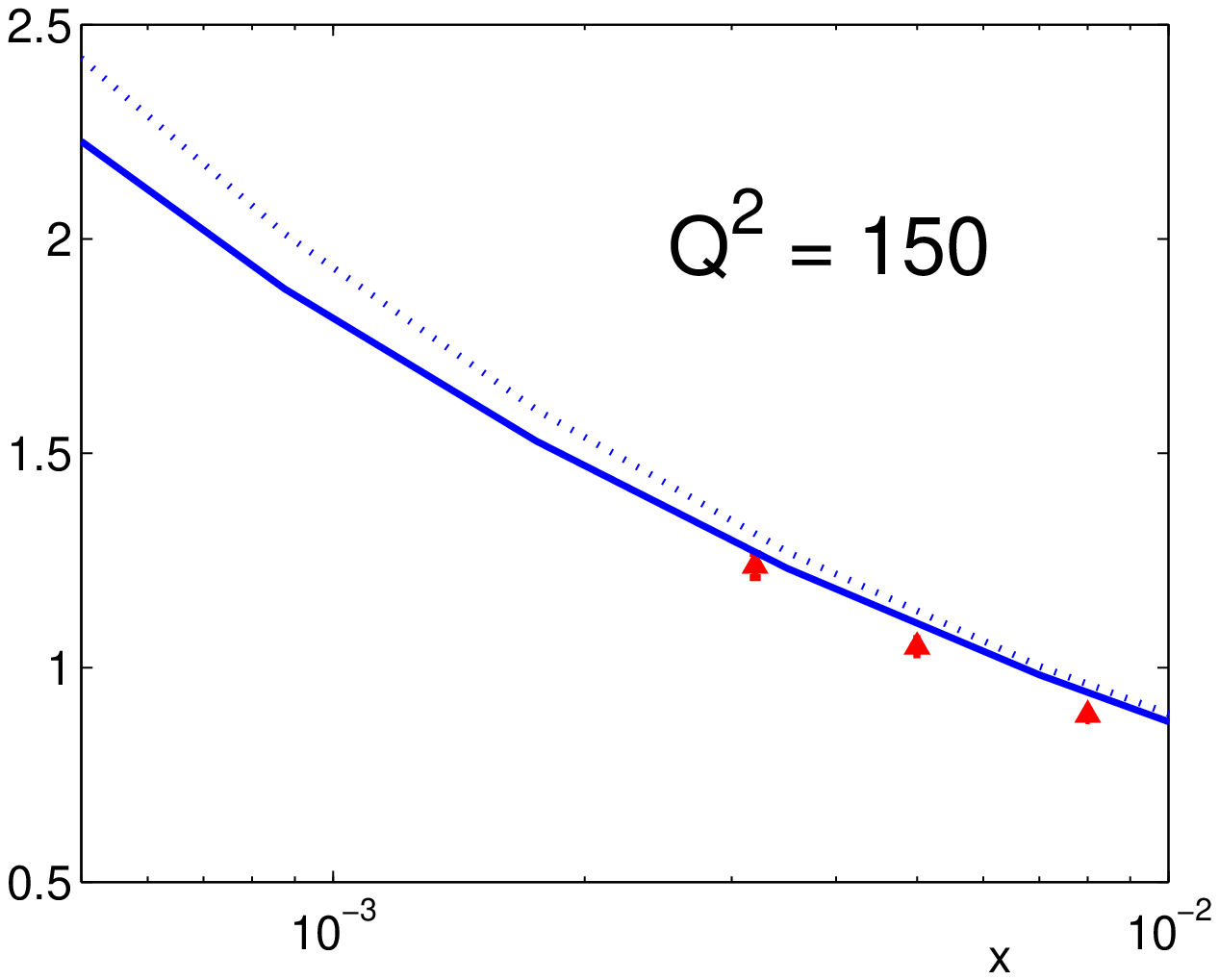,width=55mm}\\
\end{tabular}
\caption{Description of the DIS experimental data in our models.
Dotted line describes the model B while solid line corresponds to
Model A. \label{dis}} }

\section{Cross sections of hadron-hadron interaction at high energy}
\subsection{General approach}

\eq{MODN} gives a smooth continuation to the long distance physics
describing the DIS data at very low values of the photon
virtualities. We wish to extend this description to the
hadron-hadron interaction without assuming something in addition,
for example, the existence of the soft Pomeron. The simplest formula
we can write for the total cross section, is a straightforward
generalization of the formula for the DIS cross section with the
replacement of $\Psi_{\gamma^*}(r) \to \Psi_{hadron}( \{r_i\})$ ,
namely \beq \label{TXS} \sigma_{tot}\,\,=\,\, 2\,\int\,d^2 b \,\int
\prod^{n}_{i=1}\,d^2 r_i \,\sum^{n}_{i=1}\,\,|\Psi_{hadron}(
\{r_i\})|^2 \,\,N(x, r_i;b ) \eeq where $n$ is the number of dipoles
that we need to introduce to describe a hadron. For example, for a
meson we need only one dipole, while for the proton we have to introduce
at least two colorless dipoles.

The total elastic cross section can be written in the form: \beq
\label{ELXS}\sigma_{el}\,\,=\,\,\int\,d^2 b \Lb  \int
\prod^{n}_{i=1}\,d^2 r_i \,\sum^{n}_{i=1}\,\,|\Psi_{hadron}(
\{r_i\})|^2 \,\,N(x, r_i; b )\Rb^2 \eeq

For the differential elastic cross section we have the following
expression \beq \label{DELXS} \frac{d \sigma_{el}}{dt}\,\,=\,\,\Lb
\frac{1}{(2 \pi)^2}\, \int J_0(q b) d^2 b   \int
\prod^{n}_{i=1}\,d^2 r_i \,\sum^{n}_{i=1}\,\,|\Psi_{hadron}(
\{r_i\})|^2 \,\,N(x, r_i ,b)\Rb^2 \eeq where $J_0$ is the Bessel
function. It is even easier to calculate the slope in $t$ at $t=0$.
It is equal to \beq \label{SLP} B_{el}\,\,=\,\,\frac{d \ln d
\sigma_{el}/dt}{d t}|_{t=0}\,\,=\,\,\h\,\frac{\int\, b^2\,d^2
b\,\int \prod^{n}_{i=1}\,d^2 r_i \,\sum^{n}_{i=1}\,\,|\Psi_{hadron}(
\{r_i\})|^2 \,\,N(x, r_i;b)}{
 \int\,d^2 b\,\int \prod^{n}_{i=1}\,d^2 r_i \,\sum^{n}_{i=1}\,\,|\Psi_{hadron}( \{r_i\})|^2 \,\,N(x, r_i;b )}
\eeq
The process of diffractive dissociation is a more complicated
phenomenon. Indeed, the eikonal type formula of \eq{MODN} leads to
diffractive dissociation in the state of $n$ free dipoles which,
being a system with a limited number of dipoles,  cannot create a
system of hadrons  with large mass. In other words, this diffractive
production falls down as a function of produced mass. Therefore, we
can calculate  diffraction  in the region of the small mass using
the Good-Walker formula \cite{GOWA} , namely \bea \label{SDXSSM}
&&\sigma^{low\;M}_{diff}\,\,\,=\,\,\,\\
&&\int\,d^2 b\,\int \prod^{n}_{i=1}\,d^2 r_i
\,\sum^{n}_{i=1}\,\,|\Psi_{hadron}( \{r_i\})|^2 \,\,N^2(x, r_i;b
)\,\,-\,\,\int\,d^2 b \Lb  \int \prod^{n}_{i=1}\,d^2 r_i
\,\sum^{n}_{i=1}\,\,|\Psi_{hadron}( \{r_i\})|^2 \,\,N(x, r_i; b
)\Rb^2 \nonumber \eea the first term is the cross section for the
production of the system of $n$ free dipoles, while the second is the
elastic cross section. We  subtracted this term to find the cross
section of the hadron state, which is different from the initial one.
\FIGURE[h]{
\centerline{\epsfig{file=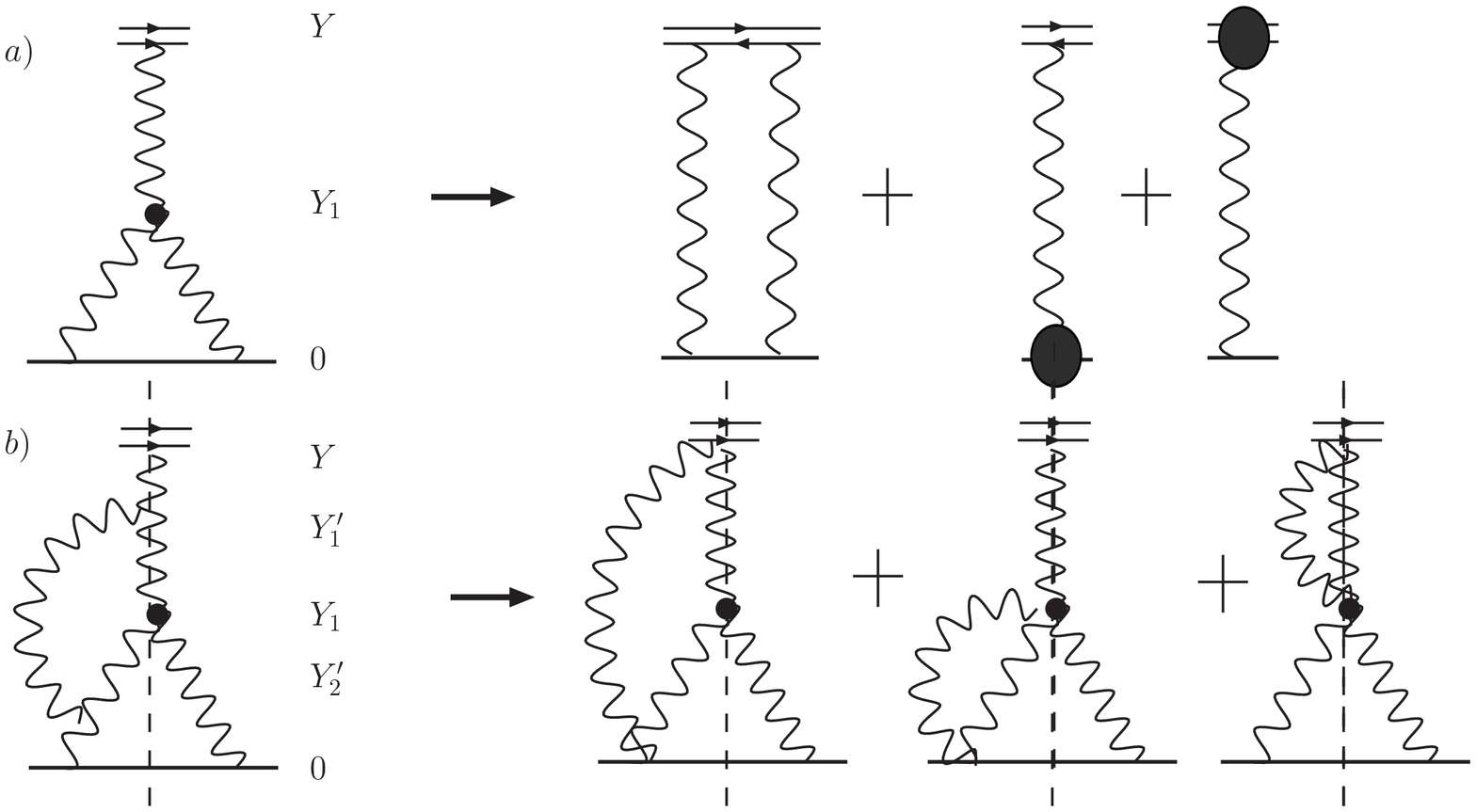,width=140mm,height=60mm}}
\caption{ Simple Pomeron diagrams for the total cross section
(\fig{pomdi}-a) and for the diffractive production in the region of
large mass ($Y-Y_1 = \ln(M^2/s_0)$, \fig{pomdi}-b). The dashed line
shows the cut Pomeron.} \label{pomdi}}. For large mass diffraction,
we have to develop a new approach. The  simple form of the
generating functional of \eq{EIKZ} stems from the fact that the
system of interacting Pomerons can be reduced to the exchange of
non-interacting Pomerons after integration over rapidity $Y_1$,   as
it is shown in \fig{pomdi}-a. In the case of diffractive production
we can also calculate the Pomeron diagrams with one cut Pomeron
(shown by dashed line in \fig{pomdi}-b) but the value of $Y - Y_1$
is fixed, namely,  $Y - Y_1 =\ln(M^2/s_0)$ where $M$ is the mass of the
diffractively produced system and $s_0 $ is the energy scale  ($s_0
\approx 1 GeV^2$). In more complicated diagrams, we have to integrate
over rapidities  (for example, over $Y'_1$ and $Y'_2$ in
\fig{pomdi}-b). For the Pomeron with the intercept larger than 1, as in
our case,  these integrations result in three diagrams in
\fig{pomdi}-b. It should be stressed that the integral over the produced
mass gives the dominant contribution at $Y -Y_1 =
\ln(M^2/s_0)\,\approx 1/\bas$. However, it has not been taken into
account in our eikonal - type model, which  only describes  the G-W
contribution to the region of small mass.

Therefore, for $d \sigma/dM^2$ at large $M^2 \gg s_0$,  for one
dipole $(x,y)$ we can write a generic formula (see Refs. \cite{DIF})
\bea \label{SDXSLM} && M^2 \,\frac{d \sigma^{high\;M}_{diff}}{ d
M^2}\,\,\,=
\,\,\,V_{1 \to 2}\Lb (x',y') \to (x',z) + (z,y') \Rb \,\,\bigotimes \,\,\,\,\exp\Lb - \Omega(s; x, y)\Rb \\
&&\Lb 1 - \exp\Lb - \tilde{\Omega}(M^2;x,y; x',y')\Rb\Rb\, \left[
\exp\Lb - \h\left\{\Omega( s/M^2; x',z) +  \Omega( s/M^2;
z,y')\right\}\Rb \,-\,\exp\Lb - \h \Omega( s; x', y')\Rb \right]^2
\nonumber \eea where $\bigotimes$ denotes all needed integrations,
and $\tilde{\Omega}$ is a new opacity for the interaction of two
dipoles: $(x,y)$ and $(x',y')$, which we can build using the same
approach as in \eq{OMEGA}. We start to analyze \eq{SDXSLM}
considering the emission of an extra gluon (see \fig{gem1}).

\DOUBLEFIGURE[h]{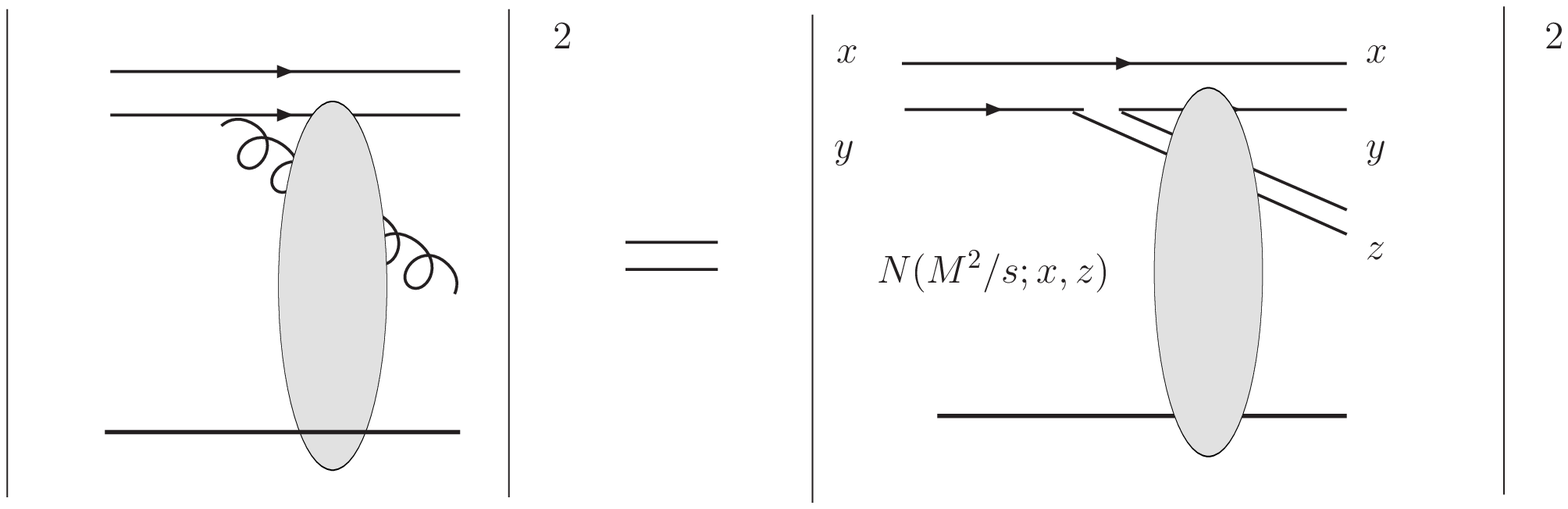,width=90mm,height=32mm}{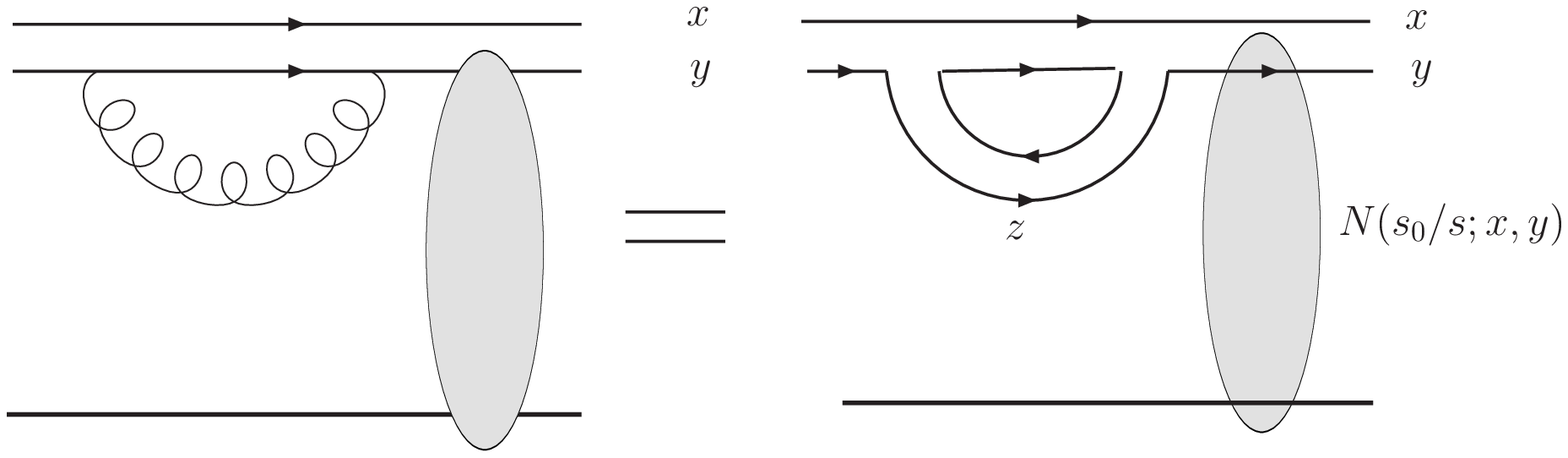,width=80mm,height=30mm}
{The  process of  diffractive production in the region of large mass
in perturbative QCD. \label{gem1}}{Feedback to the process of
elastic scattering of dipole$(x,y)$ due to the emission of an extra
gluon.\label{gem2}}

One can see that our process has three well separated stages. The
first one is a penetration of the dipole $(x,y)$ through the target
without inelastic interaction. We introduce the factor "$\exp\Lb -
\Omega\Rb$" to describe this stage. This factor sums all Pomeron
diagrams with Pomerons that carry the total energy of the process(
rapidity $Y$). See the  second diagram in \fig{pomdi}-b. The second
stage is the emission of one extra gluon. The dipole decay is
responsible  for this stage with the probability given by \eq{V}.
The last, third stage is the interaction of two  produced dipoles
$(x',z)$ and $z,y')$  with the target. This stage is taken into
account in \eq{SDXSLM} by the factor in the brackets. This factor has
been discussed in Ref. \cite{KOTU}.

In the first diagrams (see \fig{gem1}) the  second stage is very
simple: it is just the perturbative emission of one extra gluon. The
simplified formula for this diagram has the same structure as
\eq{SDXSLM}, but with a simple expression for $\exp\Lb - \Omega(s; x,
y)\Rb \Lb 1 - \exp\Lb - \tilde{\Omega}(M^2;x,y; x',y')\Rb \Rb$,
namely

~

\bea \label{SDXS1G} &&\sigma^{high\;M}_{diff}(
\fig{gem1})\,\,=\,\,\frac{ \bas\,n}{2}\,\,\int^{\infty}_{M^2_0}
\,\frac{d M^2}{M^2}\,\,\,\int d^2 b \int \prod^{n}_{i=1} \,d r^2_i
\,\,\exp\Lb -\sum^{n}_{i=2} \Omega(s/s_0; r_i,b)\Rb \,
|\Psi_{hadron}( \{r_i\})|^2\,\nonumber \\
&& \left\{\exp\Lb - \Omega(s/s_0; r_1,b)\Rb \,\,r^2_1
\int^{\infty}_{r^2_1}\frac{d r^2}{r^4}\,  \Lb 1 -   \exp\Lb - \left\{ \Omega(s/M^2;r; b)  \, -\,\h \Omega (s/M^2, r_1;b)\right\}\Rb \Rb^2 \,\,\right.\notag\\
&&\left. - \,\,r^2_1\,\int \frac{d^2 r}{2\pi\,r^2\,(\vec{r}_1 -
\vec{r})^2}\,\Lb 1\,\,-\,\,\exp\Lb - \Omega(s/s_0; r_1,b)\Rb\Rb^2
\right\} \eea where $n$ is the number of dipoles in a proton and $
\vec{r} = \vec{x} - \vec{z}$ . In \eq{SDXS1G} we consider the region
of integration where $r_i \ll r$.  The second term in curly brackets
takes into account a change for elastic scattering of the dipole
$r_1$ due to the emission of one extra gluon (see \fig{gem2}.)
\eq{SDXS1G} is written in the leading log approximation of perturbative
QCD in which we consider $\bas \ln(M^2/s_0)\,\approx\,1$ while $\as
\ll 1$.  Factor $\Lb 1 - \exp\Lb - \Omega(s/M^2;r) + \h \Omega
(s/M^2, r_l;b)\Rb\Rb $ is the amplitude of  gluon-dipole scattering
that has been discussed in Ref. \cite{KOTU}. The formulae for
diffraction production is well known (see Refs. \cite{DIF} for
details). In the same approximation \eq{SDXSLM} has the form \bea
\label{SDLM} && M^2 \,\frac{d
\sigma^{high\;M}_{diff}}{dM^2}\,\,\,=\,\,\,\frac{\bas\,n}{2}\,\,\,\int\,d^2
b \,\int \prod^{n}_{i=1} \,d^2 r_i \,\,\,
|\Psi_{hadron}( \{r_i\})|^2\,\Lb \exp\Lb - \Omega(s/s_0; r_i,b)\Rb \Rb \,\\
&& \int \,d^2 r\,\,r^2\,\,\int d^2 b' \,\,\,
 \Lb 1 - \exp \Lb -  \Omega^{BFKL}\Lb M^2/s_0; r_i, r\Rb \Rb \Rb\,\,
 \int^{\infty}_{r^2_i}\frac{d r'^2}{r'^4}\,\,\,
\Lb 1 - \exp\Lb -  \Omega(s/M^2;r';b)\Rb \Rb^2, \nonumber
\eea
where $\Omega^{BFKL}\Lb M^2/s_0; r_i, r\Rb$ is the solution of the BFKL equation with the initial condition:
\beq \label{ICOBFKL}
\frac{d\,\, \Omega^{BFKL}\Lb M^2/s_0=1; r_i, r\Rb}{ d \ln(M^2/s_0)}  \,\,\,=\,\,\,\frac{\bas}{2}\,\delta(\vec{r} - \vec{r}')
\eeq

This $\Omega$ is equal to \cite{BFKL}
\bea \label{OMEGABFKL}
\Omega^{BFKL}\Lb M^2/s_0=1; r_i, r\Rb\,\,&=&\,\,\int  \frac{d \gamma}{2 \pi \,i\,\omega(\gamma)}\,e^{ \omega(\gamma)\ln(M^2/s_0)\,\,+ \gamma \ln (r^2_i/r^2)}\,\, \\
&=& \,\,\sqrt{\frac{r_i^2}{r^2}}\,\frac{1}{\omega(0)}\,\sqrt{\frac{2
\pi}{\omega"(0) \ln(M^2/s_0)}}
\,\,e^{\omega(0)\ln(M^2/s_0)\,\,-\,\,\frac{\ln^2(r^2/r^2_i)}{2
\omega"(0)\,\ln(M^2/s_0)}} \nonumber \eea where the eigenvalues of
the BFKL equation $\omega(\nu)$ can be found in Ref. \cite{BFKL}.

The factor $1 - \exp\Lb - \Omega^{BFKL}\Rb$ describes the inelastic
cross section which in terms of the Pomeron diagrams of \fig{pomdi},
reflects the possibility that in the fourth of \fig{pomdi}-b, we can
have two cut upper Pomerons.

We would like to stress again that \eq{OMEGABFKL}  describes the
dependence  of the diffractive production cross section, in the
region of large mass, while \eq{SDXSSM} is written for the  low mass
diffractive cross section. Therefore, to calculate the cross section
for diffractive production, we need to calculate \beq \label{SMDIFF}
\sigma_{diff}\,\,\,=\,\,\sigma^{low\;M}_{diff}\Lb \eq{SDXSSM}
\Rb\,\,+\,\,\sigma^{high\;M}_{diff}(\eq{SDLM}) \eeq

\subsection{Hadronic  wave functions}
As seen from the formulae in the previous subsection, we need to know
the hadronic wave functions. We have made an assumption in writing
these formulae, that the correct degrees of freedom  at long
distances  are the colorless dipoles at least at high
energy\cite{DFK,BATAV}. This is one of the strongest assumptions in
our approach. It is enough to recall that for a long time the
constituent quarks have been considered as a good candidate for the
correct degrees of freedom in the entire range of energy.  The only
support for such an approach, can be seeen in the success of the Heildelberg
group (see Refs. \cite{DFK} and references therein), in the description
of soft interactions using this ansatz.

In the case of mesons, we have only one colorless dipole and we take
the transverse wave function in the form of a simple Gaussian,
namely \beq \label{MPSI} |\Psi_{meson} \Lb r \Rb
|^2\,\,\,=\,\,\frac{1}{\pi S^2_M}\,e^{ - \frac{r^2}{S^2_M}} \eeq
where $S_M$  is a parameter that can be found from the
electromagnetic radii, namely, $R_\pi = 0.66 \pm 0.01 fm$ and $R_K =
0.58 \pm 0.04 fm$ \cite{ER}. Using $S_M = \sqrt{\frac{8}{3}} R_M$ we
obtain $S_\pi = 1.08 fm$ and $S_K = 0.95 fm$.
\FIGURE[h]{\begin{minipage}{90mm}
{\centerline{\epsfig{file=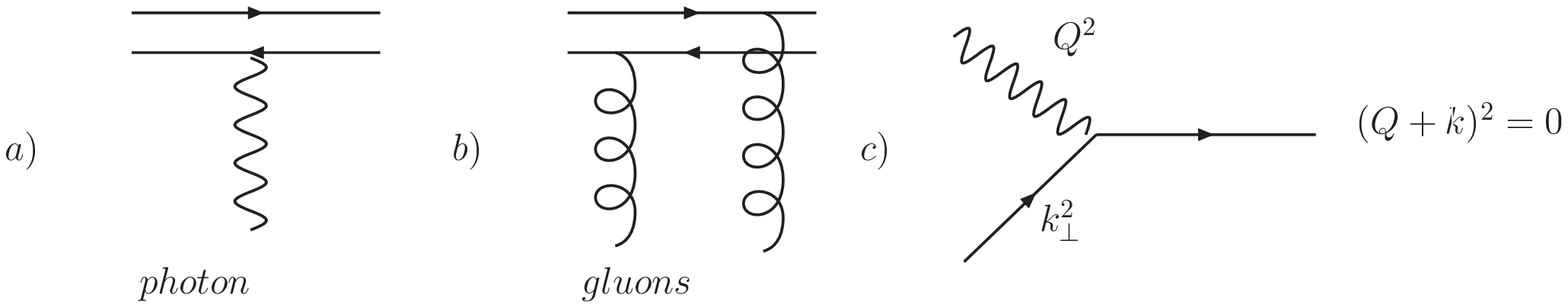,width=85mm}}
\caption{Electromagnetic (\fig{fact}-a) and two gluon (\fig{fact}-b)
form factors. for a meson. \fig{fact}-c shows the interaction of a
virtual photon with a parton.} \label{fact} }
\end{minipage}
}

However, electromagnetic form factors gives us the space distribution
of the electric charge inside of the hadron,  while in our case the
distribution of the density of partons (dipoles) is probed by the two
gluon interaction (see \fig{fact}-b). This distribution could be
different from the charged one. We prefer to find the value of $S_M$
from the experimental data and, therefore, $S_M$ as well as $S_p$
(see below), will be the only fitting parameters in our
description of the soft experimental data.

For the baryon, the situation is more complicated: we can have two or even
three dipoles.  Follow Ref. \cite{DFK}, we choose the proton wave
function in the simple form for the two dipole model \beq \label{BPSI}
|\Psi_{proton}(r_1,r_2)|^2\,\,=\,\,\frac{1}{(\pi\,S_p)^2}\,\,e^{-
\frac{r^2_1 + r^2_2}{S^2_p}} \eeq where two dipoles are defined as
$\vec{r}_1 = \vec{R}_1 - \vec{R}_2$ and $ \vec{r}_2 = \vec{R}_3 - \h
\Lb \vec{R}_1 + \vec{R}_2 \Rb$ where $\vec{R}_i$ is the position of
the  constituent quark $i$. In \eq{BPSI} $S_P =
\sqrt{\frac{3}{2}}\,R_p \,=\,1.05 fm$ for $R_p = 0.862 \pm 0.012
fm$\cite{ER}.

However, for large $N_c$ it is proven that the baryon consists of
$N_c$ dipoles \cite{LNCB}. Therefore, we consider the alternative
assumption for $\Psi_{proton}(r_1,r_2,r_3)$, namely

\beq \label{BPSI1}
|\Psi_{proton}(r_1,r_2,r_3)|^2\,\,=\,\,\frac{1}{(\pi\,S_p)^3}\,\,e^{-
\frac{r^2_1 + r^2_2\,+\,r^2_3}{S^2_p}} \eeq where dipoles have the
size $\vec{R}_i - \frac{1}{3} ( \vec{R}_1 + \vec{R}_2 + \vec{R}_3)$.
In this case, from the electromagnetic radius of the proton, follows the
value of $S_p = R_p =0.862 \pm 0.012 fm$.

In \eq{BPSI}, we take the size of two dipoles to be equal. We did
this for simplicity, since even in the constituent quark model (CQM),
$<|r^2_2|> = 4/3 <|r^2_1|>$.  Our hope is that the hadron interaction
will be determined by the behavior of  the  dipole  amplitude in the
saturation domain, where the sensitivity to the size of dipoles
is expected to be weak. On the other hand,  the experimental ratio
$\sigma_{tot}(\pi -p)/\sigma_{tot}(p-p)\,\approx\,2/3$ in the CQM stems
from quark counting. In our approach, the dipole counting leads to
the value of this ration $1/2$ (for the two dipole model for a proton)
or even 1/3 (in the three dipole model)  which contradicts  the
experimental data. The only way to obtain a reasonable description,
is to hope that the perturbative QCD dependence of  the dipole cross
section on the size of the  dipoles $\sigma \propto r^2$ will remain in
the entire accessible range of energies. Fortunately, Ref.
\cite{BATAV} demonstrates that this is the case, and this
result encourages us to search for the description of the soft
processes using the dipole hypothesis for the hadronic wave
functions.

\subsection{Energy variable in the model}

In section 2.2 and 2.3 we have discussed our approach introducing
the typical energy variable for deep inelastic scattering:
$x_{Bjorken} \equiv x = Q^2/s$ where $Q^2$ is the photon virtuality
and $s$ is the energy. However, with this variable we cannot discuss
the soft processes which have $Q^2 =0$.  We reconsider the
derivation of the Bjorken variable to introduce a new energy
variable which will have a limit $x_{soft} \to x$ at $Q^2 \gg \mu^2$
where $\mu$ is the scale for the soft processes. From \fig{fact}-c
we have a relation
\begin{align} \label{NEWX}
(q + k)^2& =  - Q^2 + x_{soft} s \,\,+ \,\,k^2\,\,=\,\,0; &  x_{soft} & = \Lb Q^2 + k^2_{\perp}\Rb/s
\end{align}
where $q$ is the momentum of virtual photon ($q^2 - Q^2$).

In \eq{NEWX} we used the fact that at high energy, $k^2 = -k^2_{\perp}$. Since in the saturation domain $ k^2_{\perp}\,\,=\,\,Q^2_s (x)$, we obtain the final expression for our energy variable
\begin{align} \label{NEWX1}
 x_{soft}&\,\,=\,\,\frac{Q^2 + Q^2_s\Lb x_{soft}\Rb}{s}\,\,\xrightarrow{Q^2 \rightarrow 0}\,\,\frac{Q^2_s\Lb x_{soft}\Rb}{s}; &  x_{soft}&\,\,=\,\,\frac{Q^2 + Q^2_s\Lb x_{soft}\Rb}{s}\,\,\xrightarrow{Q^2 \gg Q^2_s}\,\,x_{Bjorken};
\end{align}
We believe that one of the main advantages of the saturation
approach is the natural choice of the energy variable given by
\eq{NEWX1}, which depends on the new scale:  the saturation momentum.

The value of the saturation momentum in our model, we find by resolving
the following equation: \beq \label{QS} \Omega\Lb
x_{soft},r^2_{sat}\,b=0\Rb\,\,=\,\,1 \,\,\,\mbox{with}\,\,r^2_{sat}
= 4/Q^2_S\Lb x_{soft} \Rb \eeq

Partons (dipoles) with size $r_{sat}$ are populated densely in the
hadron disc, and $Q_s$ gives the new scale which shows that  at $Q^2
< Q^2_s$   the hard process reaches the saturation domain.

\subsection{Final formulae for proton (antiproton) - proton scattering}
Using the form of the proton  wave function given by \eq{BPSI}, we can rewrite \eq{TXS} - \eq{SLP}
in a more accurate form, taking into account the simultaneous interaction of two dipoles (see \eq{ppfin}).
\FIGURE[h]{
\centerline{\epsfig{file=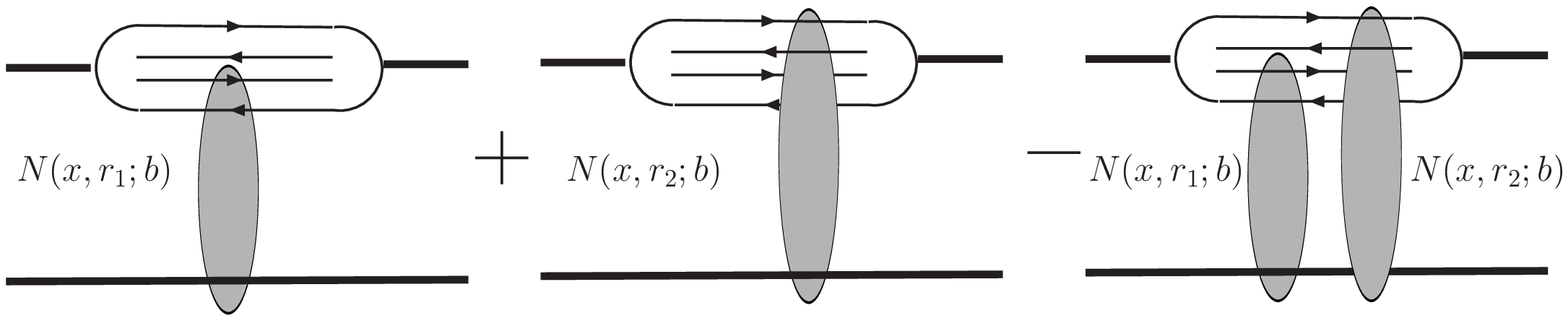,width=140mm}}
\caption{ Total cross section for proton( antiproton) -proton collision  taking into account the interaction of two dipoles.}
\label{ppfin} }
For the total cross section, the general formula has the form which follows directly from \fig{ppfin}, namely
\bea \label{TXS1}
&&\sigma_{tot}\,\,= \,\, 2\,\int\,d^2 b\,d^2 r_1\,d^2 r_2\,\,| \Psi_{proton} \Lb r_1,r_2\Rb|^2\\
&&\,\, \Lb N(x, r_1;\vec{b} - \frac{1}{3} \vec{r}_2) \,\,+\,\,N(x,
r_2;\vec{b} - \frac{1}{6} \vec{r}_2) \,\,- \,\,N(x, r_1;\vec{b}  +
\frac{1}{3} \vec{r}_2)\, N(x, r_2;\vec{b}  + \frac{1}{6} \vec{r}_2)
\Rb \nonumber \eea where $\vec{b}$ is the impact parameter or the
distance between $\frac{1}{3}( \vec{R}_1 + \vec{R}_2 + \vec{R}_3)$
and the position of the target nucleon.  One can see that we need to
replace \bea \label{REPLCE}
&&\sum^{n}_{i=1}\,\,|\Psi_{hadron}( \{r_i\})|^2 \,\,N(x, r_i;b )\,\,\longrightarrow\,\,\,|\Psi_{hadron}( r_1,r_2)|^2 \\
&&\Lb N(x, r_1;\vec{b} - \frac{1}{3} \vec{r}_2) \,\,+\,\,N(x, r_2;\vec{b} - \frac{1}{6} \vec{r}_2) \,\,- \,\,N(x, r_1;\vec{b}  +  \frac{1}{3} \vec{r}_2)\,
N(x, r_2;\vec{b}  + \frac{1}{6} \vec{r}_2) \Rb \nonumber
\eea
in all of the equations of section 3.1.

Using the explicit form of the proton wave function of \eq{BPSI}, we can
easily  rewrite the integrals in \eq{TXS} - \eq{SLP}, in the form
\bea \label{NEWF}
& &\int\,d^2b\,d^2r_1\,d^2r_2\,\,| \Psi_{proton} \Lb r_1,r_2\Rb|^2\\
& & \Lb N(x, r_1;\vec{b} - \frac{1}{3} \vec{r}_2) \,\,+\,\,N(x,
r_2;\vec{b} - \frac{1}{6} \vec{r}_2) \,\,- \,\,N(x, r_1;\vec{b}  +
\frac{1}{3} \vec{r}_2)\,
N(x, r_2;\vec{b}  + \frac{1}{6} \vec{r}_2) \Rb \nonumber \\
&=& \int dr^2_1\,db^2_1\,db^2_2\,\frac{1}{S_p^2}\,\exp\Lb - \frac{ r^2_1 + 4\,b^2_1 + 4\,b^2_2}{S^2_p}\Rb\,  I_0\Lb\frac{4}{S^2_p}b_1\,b_2\Rb \,\,\nonumber\\
&\times &  \Lb N(x, r_1;b_1)\,\,\,+\,\,\,N(x,
r_1;b_2)\,\,\,-\,\,N(x, r_1;b_1)\,\,N(x, r_1;b_2)\Rb \nonumber \eea
In \eq{NEWF}, we used the formula {\bf 8.431(3)} of Ref. \cite{RY}.

Since the typical impact parameter increases with energy, (see the next section) and the $r^2/3 $ is small (about $
1/3 \,fm$ ), we can safely neglect the shift in the definition of the impact parameters in different amplitudes, replacing
\bea \label{REPLCE1}
&&\sum^{n}_{i=1}\,\,|\Psi_{hadron}( \{r_i\})|^2 \,\,N(x, r_i;b )\,\,\longrightarrow\,\,\,\\
&& |\Psi_{hadron}( r_1,r_2)|^2 \,\,
\Lb N(x, r_1;\vec{b} ) \,\,+\,\,N(x, r_2;\vec{b} ) \,\,- \,\,N(x, r_1;\vec{b} )\,
N(x, r_2;\vec{b} ) \Rb \nonumber \\
&&|\Psi_{hadron}( r_1,r_2)|^2 \,\,\Lb 1\,\,\,-\,\,\,\exp\left[ - \h
\Lb \Omega\Lb x, r_1; \vec{b}\Rb + \Omega\Lb x, r_2; \vec{b}\Rb
\Rb\right] \Rb \label{2DM} \eea

This simplified formula works quite well in all the observables of
section 3.1, except for $d \sigma_{el}/d t$ at large values of $t$.
The formula of \eq{2DM} has a simple generalization to the case of the
three dipole proton model . In the three dipole model, for a proton the
cross section has the form: \bea \label{TXS3}
&&\sigma_{tot}\,\,= \\
&&2\,\int\,d^2 b\,d^2 r_1\,d^2 r_2\,\,| \Psi_{proton} \Lb r_1,r_2,r_3\Rb|^2
\,\, \Lb 1\,\,\,-\,\,\,\exp\left[ - \h \Lb \Omega\Lb x, r_1; \vec{b}\Rb   + \Omega\Lb x, r_2; \vec{b}\Rb
+ \Omega\Lb x, r_3; \vec{b}\Rb \Rb\right]\Rb \notag
\eea

It should be stressed that the  second term in \eq{TXS1} plays a
very important role at high energies, since it provides the observation that the
cross section approaches a black disc limit. Indeed, only with this
non-linear term in the region where the dipole amplitude tends to unity,
($N \to 1$) the factor in parenthesis ($ 2 N - N^2$) also approaches
1 leading in the black disc regime.

\subsection{Description of the soft cross sections}
In the previous sections we have built our model and fitted all
needed parameters using the DIS experimental data.  In this section,
we compare the model with the experimental data without any
additional fitting parameters.

\subsubsection{Total cross sections}
Using \eq{TXS} , \eq{MPSI} and \eq{BPSI} we calculate the cross
sections of   pion-proton, kaon -proton, proton-proton and
antiproton - proton scattering. The results are presented in
\fig{txsm} and \fig{txs}.  In addition to \eq{TXS},  we add a
contribution of the secondary Regge trajectories  using the
Donnachie and Landshoff parametrization \cite{DL}. The agreement
with the data is  good, and shows that the saturation can replace the soft
Pomeron, which has been used for fitting the soft experimental data.
However,  as has been mentioned, we still have one fitting
parameter, namely, $S_M$ and $S_p$ in the hadron wave function. The
values of $S_{\pi}$,$S_K$ and $S_p$ is given in Table 2. It is
interesting to notice that the relation between the fitted values of $S$,
is the same since it stems from the electromagnetic radii.

\TABLE[ht]{
\begin{tabular}{|c|c|c|c|} \hline
Model & $S_\pi$ (fm)   &   $S_K$ (fm) & $S_p$(fm)  \\
\hline \hline A(2D) &0.515  & 0.552 & 0.561 \\\hline
A(3D) &0.515  & 0.552 & 0.458\\
\hline B(2D) &0.556  & 0.597  & 0.615 \\ \hline B(3D) & 0.556 &
0.597 & 0.519 \\ \hline \hline
\end{tabular}
\caption{\label{S_tab}\emph{Values of parameters in hadronic wave
function that give  description of the soft data}.} }

One can see that we are able to describe both the value and the
energy behavior of the total cross section for meson-proton and
baryon - proton interaction in our two models. However,the model B
leads to a better description demonstrating, in our opinion, that
the main idea to replace the phenomenological soft Pomeron by the
behavior of the 'hard' amplitude in the saturation region,  is
fruitful . As has been discussed in the previous subsection, the most
surprising fact is that the model describes both the proton-proton
and  meson-proton cross sections.

\FIGURE[h,t]{
\begin{tabular}{c c}
\epsfig{file=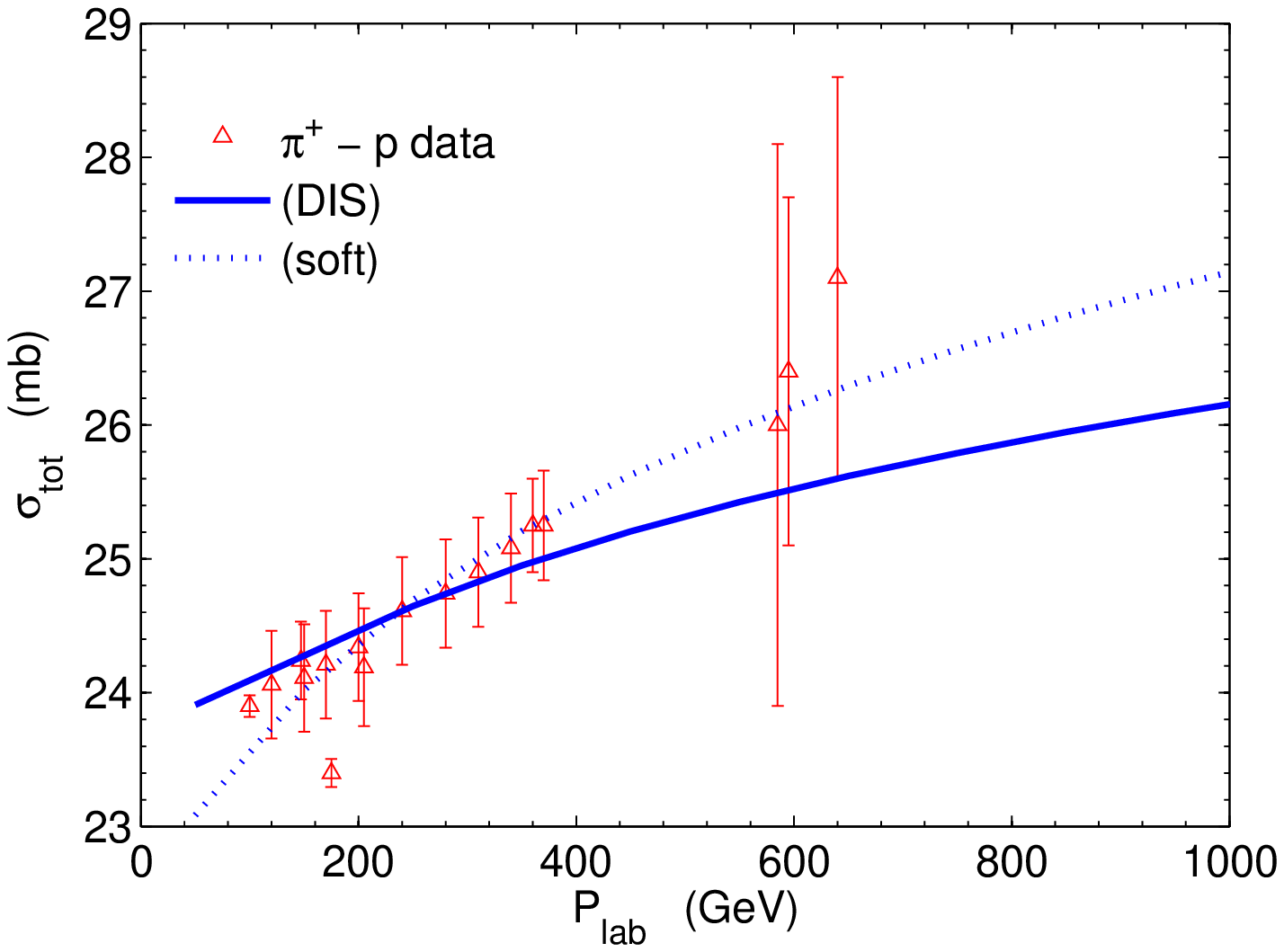,width=70mm,height=55mm} &\epsfig{file=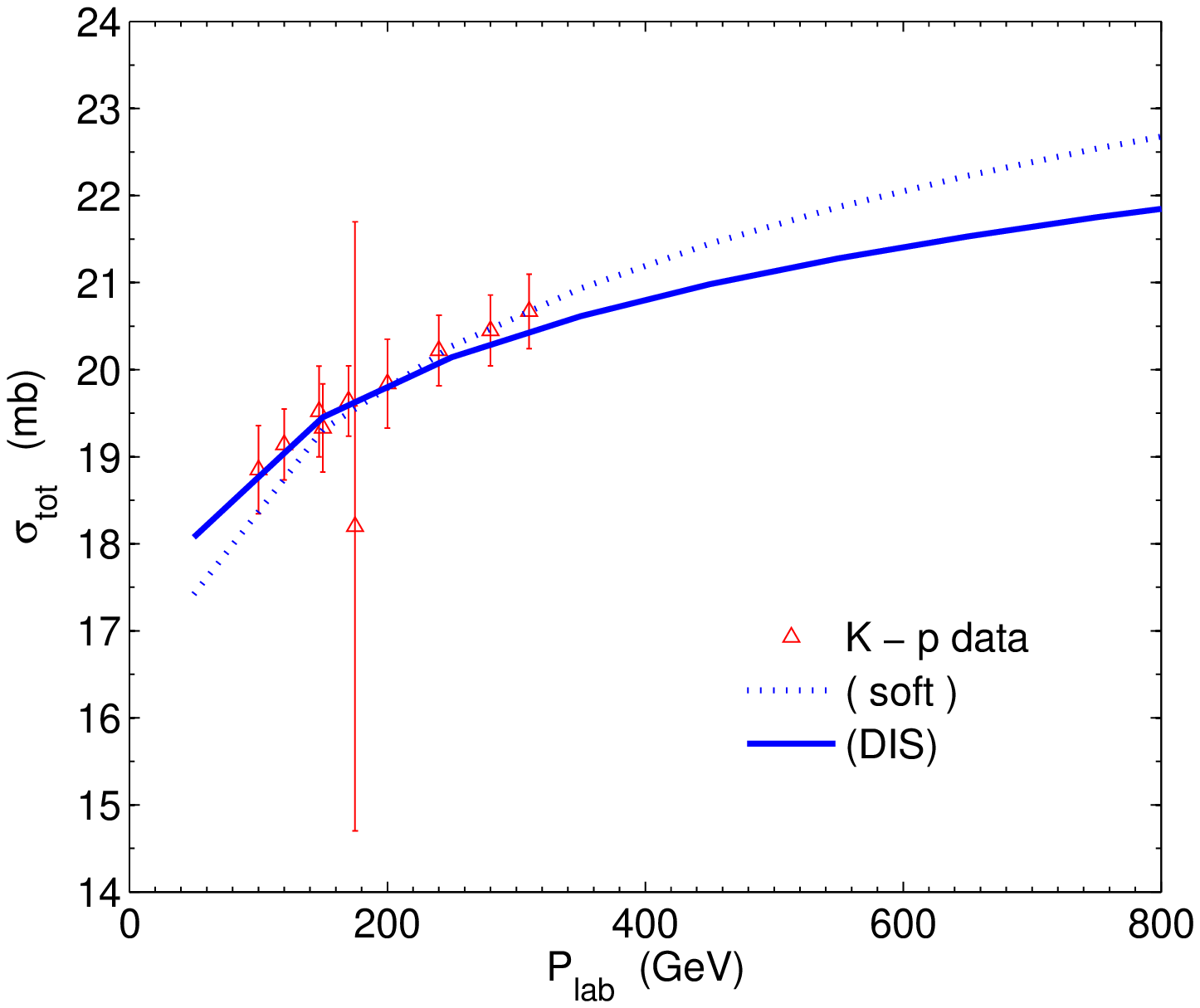,width=70mm,height=55mm}\\
\fig{txsm} - a & \fig{txsm} -b\\
\end{tabular}
\caption{ Total cross section for $\pi - p$ (\fig{txsm}-a)  and $K  -  p$ (\fig{txsm}-b)  as a function of energy.
The solid line (DIS) corresponds to the model A which describes the DIS data  with  $\chi^2/d.o.f.  \,<\, 1$, while
the dotted line  presents the model B (soft) with a worse $ \chi^2/d.o.f.  \, \approx 3$, but which leads to a better
description for DIS, and soft processes data together. The number of dipoles in the proton wave function is shown in the legend. }
\label{txsm}
}

\FIGURE[h,t]{
\centerline{\epsfig{file=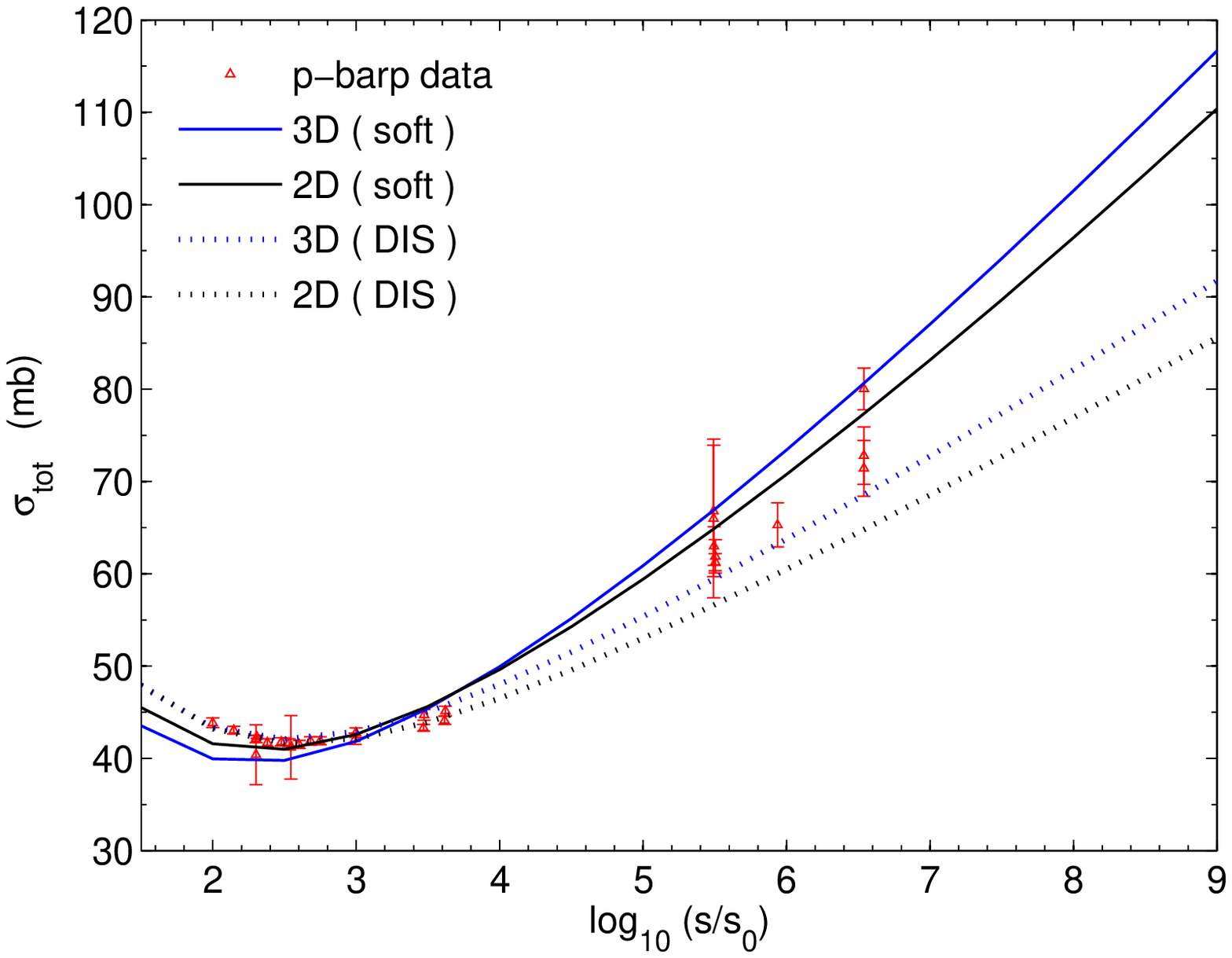,width=140mm}}
\caption{ Total cross section for $ p \bar{p}$  scattering  versus energy in our approach.
Solid lines correspond to the model A, which describes the DIS data  with  $\chi^2/d.o.f.  \,<\, 1$, while
the dotted lines present the model B (soft) with worse $ \chi^2/d.o.f.  \, \approx 3$ but which leads to a better
description for DIS, and soft processes data together.}
\label{txs} }

This fact can be explained only by the dependence of the dipole cross
section, on the size of the dipole which the model reproduces. Since such
a sensitivity can be only outside the saturation region, or close to
the boundary between the saturation domain, and the perturbative QCD
region. Therefore, we conclude that the soft data allow us to obtain
new information on this very important transition region between the
saturation and perturbative QCD region.  It should be noticed that we  also
reproduce the difference between the interaction of pions and kaons.
Since our  QCD  dipole cross section does not depend on the
mass of the quarks, this difference also  reflects the difference in the size
of the dipoles, inside pion and kaon.

\FIGURE[h,t]{ \centerline{
\epsfig{file=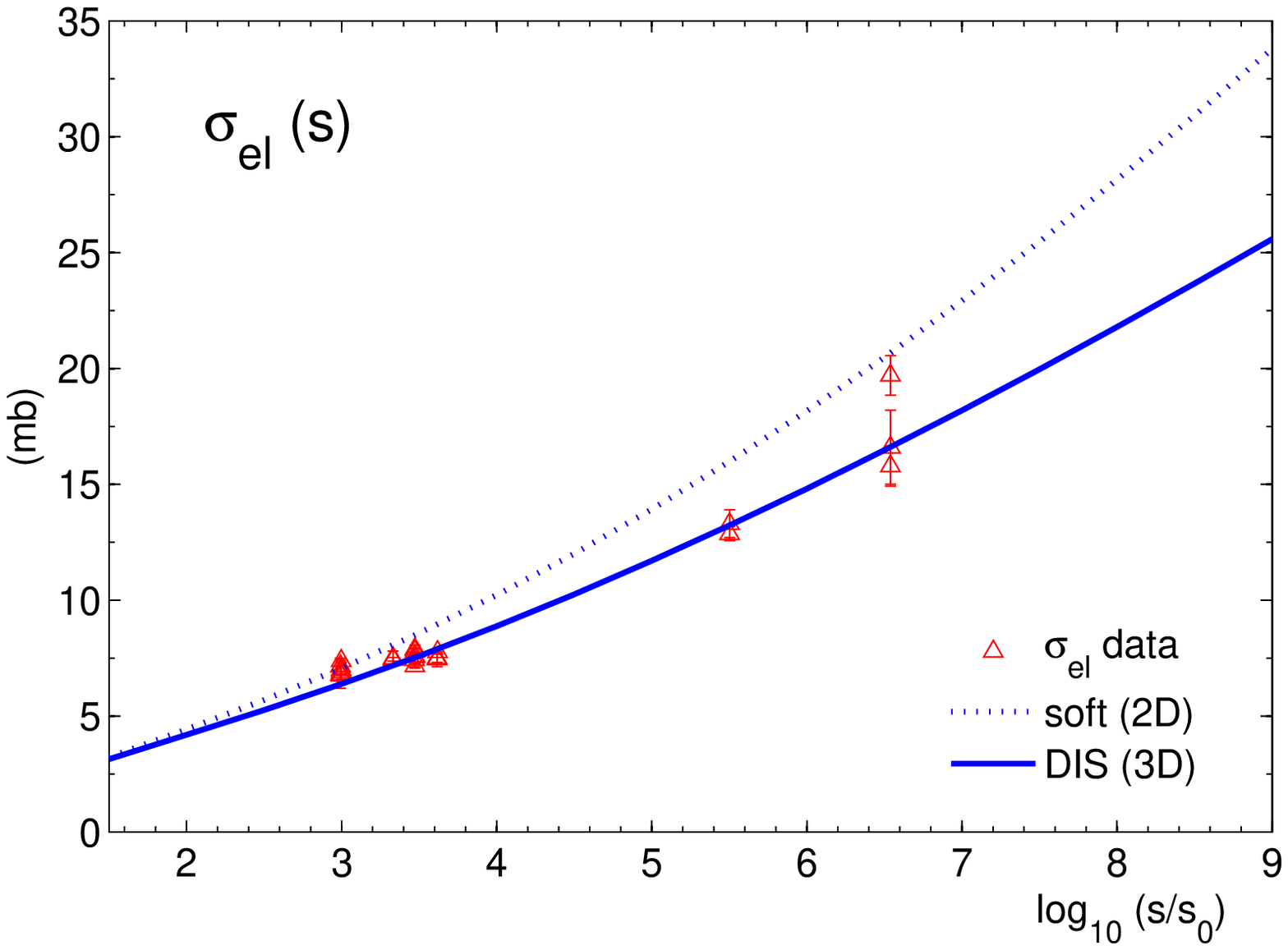,width=140mm}} \caption{ Elastic
cross section for proton-proton scattering  versus energy in our
approach. Solid line shows the comparison with the model A with 3
dipoles in the proton wave function which described DIS  with
$\chi^2/d.o.f.  \,< \, 1$ while in the dotted line we plot the
comparison with the experimental data the model B  with
$\chi^2/d.o.f.  \,\approx \, 3 $ for DIS data  for the two dipole
model of proton. } \label{xsel} }

\FIGURE[h,t]{
\centerline{
\epsfig{file=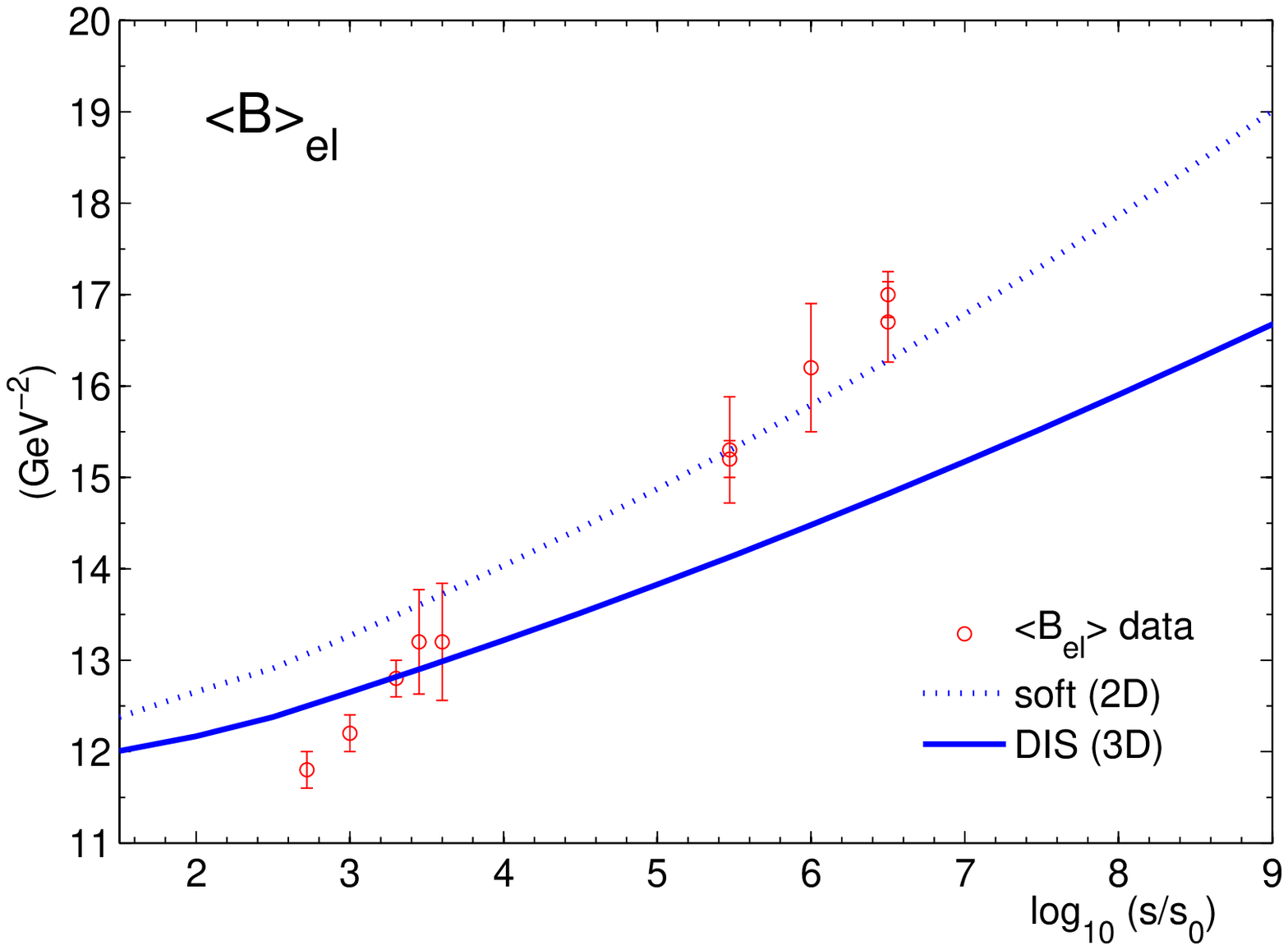,width=140mm}}
\caption{ Slope for the cross section for proton-proton scattering  versus energy in our approach. Notations are the same as in \protect\fig{xsel}. }
\label{bel} }

\subsubsection{Elastic cross section: energy dependence}

We concentrate our efforts on the proton(antiproton) interactions since
meson-proton collisions have been studied experimentally only in the
limited energy range (see \fig{txs}). The energy dependence of the
elastic cross section is presented in \fig{xsel}.  We use \eq{ELXS}
for performing our calculations. In \fig{xsel} - a we compare with
the experimental data our model B, while in \fig{xsel} - b, is plotted
our prediction for model A (three quarks). One can see that we
obtain a good agreement with the experimental data, without any
fitting parameter.

\subsubsection{Elastic  cross section: t-dependence}
Using \eq{SLP} we calculate the slope for the elastic cross section (see
\fig{bel}). One can see that in the model B, we reproduce the value and
energy behavior of the slope at high energy, but overshoot the
experimental  value of this slope at sufficiently low energy. We did
not include the secondary Reggeons, since we need a piece of information on
the slope of the contribution of  the secondary Reggeons. We also
need to take into account the fact that our  profile function $S(b)$
corresponds to a power-like form factor in the $t$ representation, and the
value of the slope for such a function is sensitive on the range of
$t$, where the slope was found experimentally. We calculate the slope
at $t =0$, while the slope at an average value of $|t| = |t_0|> 0$ is
less than at $t=0$. Therefore, it is better to compare with the
experimental data the $t$ dependence of the differential elastic
cross section, given by \eq{DELXS}. Such a comparison, one can see in
\fig{dsigdt}, where the data at small $t$ is described by our model
even at low energies ($\sqrt{s} = 23.5 GeV\,\, \mbox{and} \,\,62.5
GeV$). Notice that at such low energies,  the slope, that we
predict, exceeds the experimental one (see \fig{bel}). However, in
model A, the slope is too small and, actually, this is the key
difference between the two approaches. The origin for such a difference in
the framework of our approach  is clear: a different energy
dependence of the dipole-proton amplitude in the transition region
between  the saturation domain, and the perturbative QCD region.
Indeed, \fig{dis} shows that model B has a steeper behavior in this
region in comparison with model A.

In \fig{bel} we compare our prediction with the standard
phenomenological parametrization for the slope in the soft Pomeron
model:   $B_{el} = B_0 + 2 \alpha' \ln(s/s_0)$ with $\alpha' = 0.25
\,GeV^{-2}$ \cite{DL}  and $B_0 = 9 \,GeV^{-2}$. One can see that
our prediction in model B is in perfect agreement with this model
at  high energies, while we predict a large slope at low energies.
However, model A leads to a smaller slope than is seen in
experiment. It should be noticed, the comparison with the
experimental data (see \fig{dsigdt}) shows that model A fits the
experimental data.

\FIGURE[h]{
\centerline{\epsfig{file=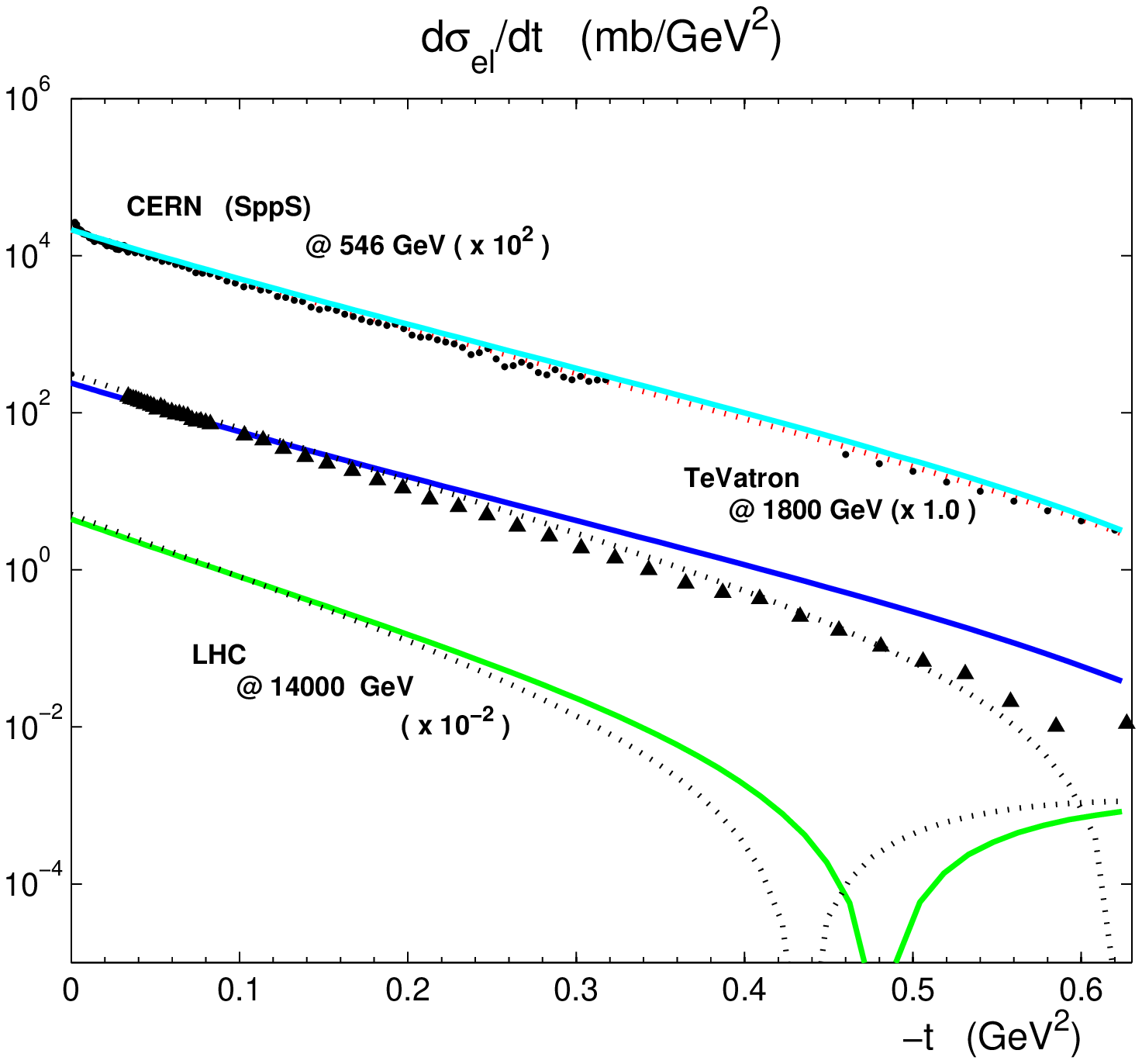,width=130mm,,height=110mm}}
\caption{ $t$-dependence of the elastic cross section for different
energies at  $t \,\leq\, 0.6 GeV^2$. The solid line corresponds to Model
A predictions, while the dotted line describes Model B behavior of the
differential cross section. Both models reproduce small $t$ behavior
of the experimental data in spite of the poor description of the
elastic slope in model A. \label{dsigdt} }}

In \fig{dsigdt} we plot the $t$-dependence for the different energy.  We
would like to draw your attention to two interesting features of our
model B: it describes the experimental data quite well at small
values of ${t} \leq 0.6\, GeV^2$; and it predicts the minimum at the
LHC energy at rather small values of $t$ ($|t|  \approx 0.45
\,GeV^2$). The appearance of this minimum shows that our model
reproduces the effective shrinkage of the diffractive peak which
results in moving the typical diffractive minima in the region of
smaller $t$, in comparison with lower energies.  Since we did not
include the real part of our scattering amplitude, we see a deep
minimum. However, the real part of the amplitude will make our
minima more shallow. Model \emph{A} leads to a smaller value of
$B_{el}$, which actually leads to a dip at large values of $t$, as compared to model \emph{B}. For our rather crude models, such a description
looks successful and , we hope, will encourage others to look seriously into
attempts involving the description of the experimental, data based on the
saturation regime of QCD.

\subsubsection{Diffraction production: energy behavior of the cross section}

\FIGURE[ht]{\centerline{\epsfig{file=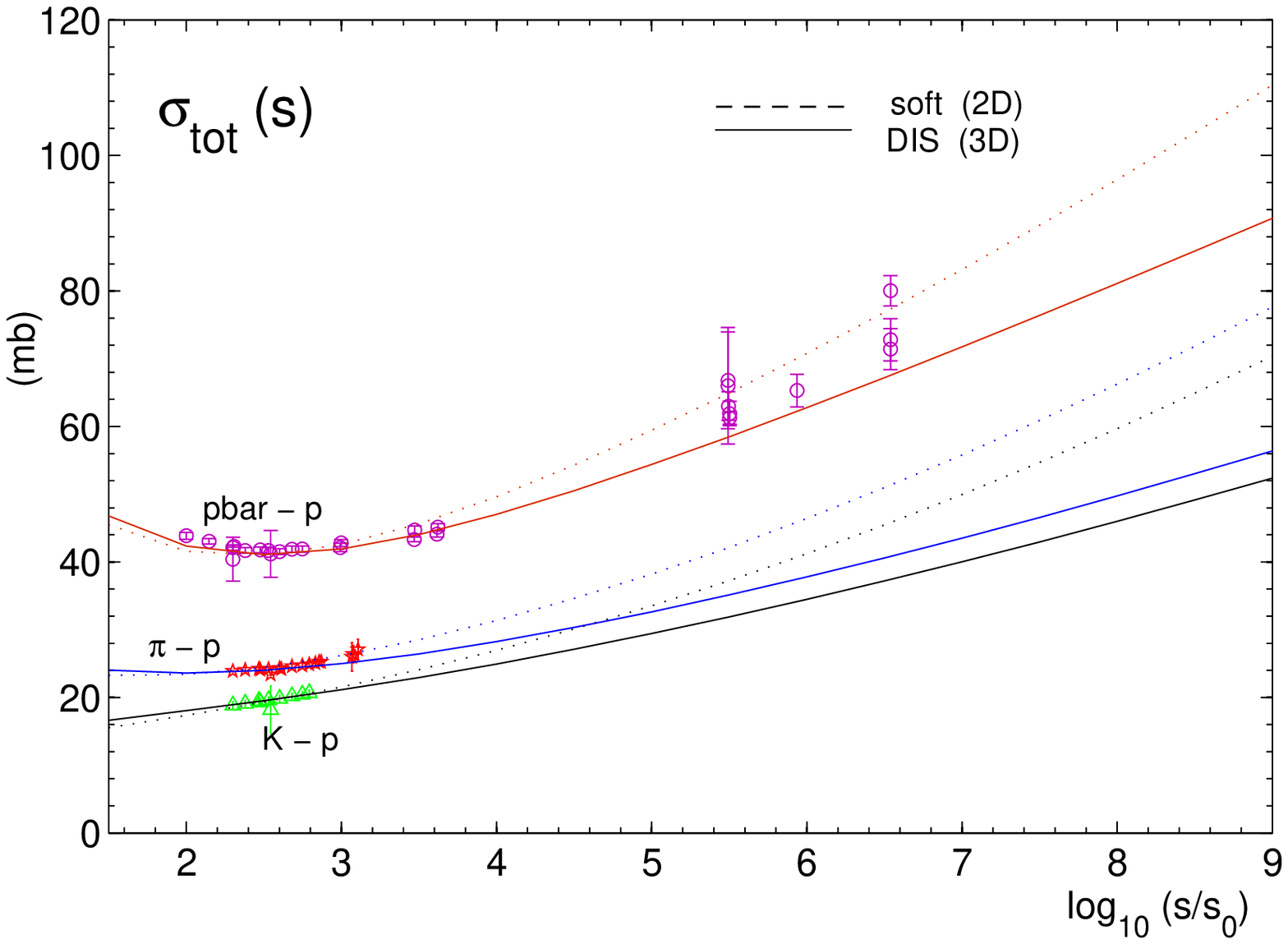,width=140mm}}
\caption{Total cross section for meson-proton and
proton(antiproton)-proton scattering scattering at high energies
(extrapolation). Notations are the same as in \protect\fig{xsel}.}
 \label{mespro}}

In \fig{sdxs} the total cross section (integrated over the masses of
the produced particles) of diffractive production is plotted. We use
\eq{SDXSLM} to calculate this cross section. However, we multiply
this formula by 2 since the two protons: projectile and target, can
dissociate diffractively. One can see from this picture that we
reproduce values and energy dependence in agreement with the
experimental data. The qualitative behavior of the experimental data
on  $\sigma_{diff}$ versus energy is characterized by the slow
dependence on energy in the region of high energies. This fact is
perfectly reproduced by our model.
\FIGURE[ht]{\centerline{\epsfig{file=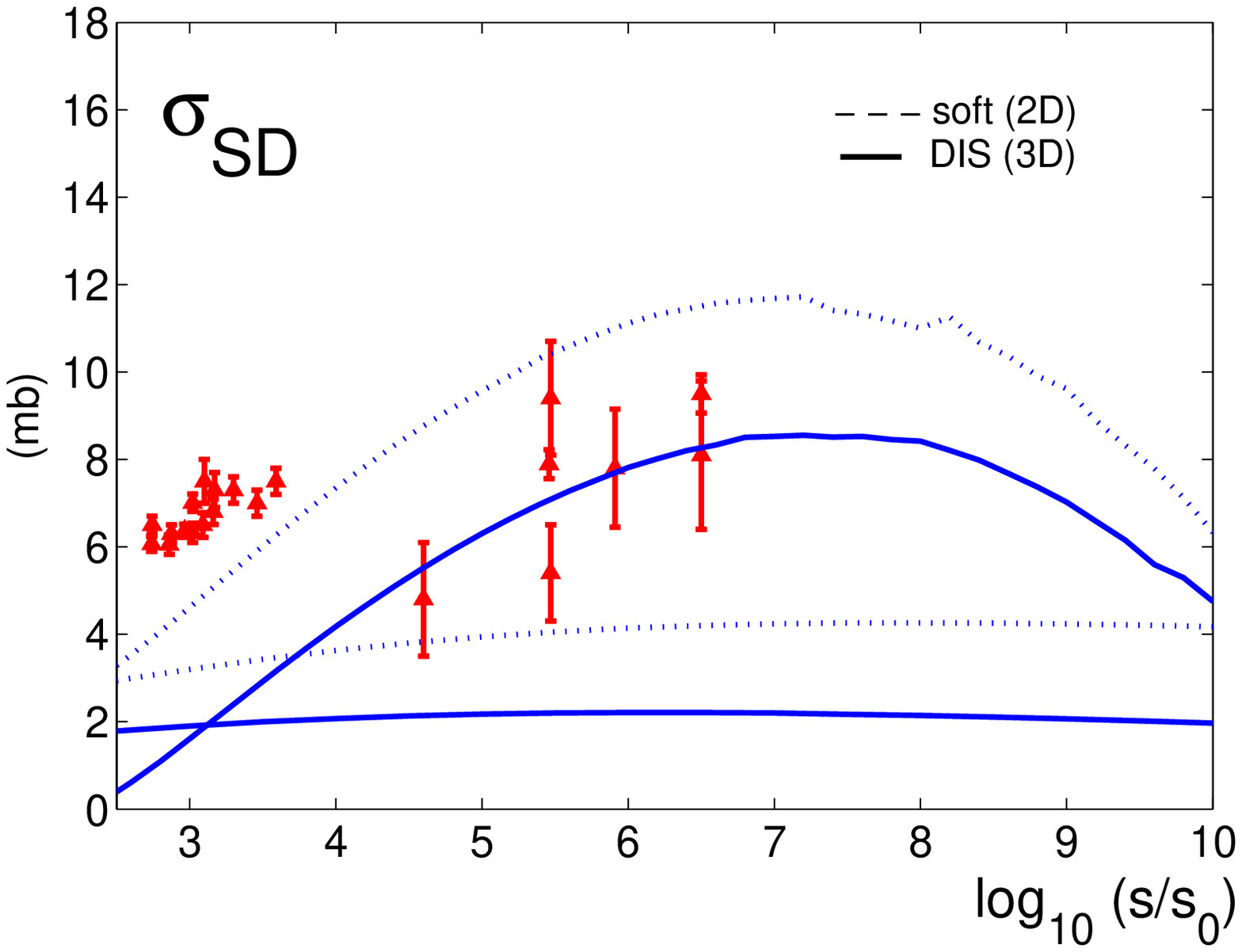,width=140mm}}
\caption{ Cross section for single diffraction production for
proton(antiproton)-proton scattering versus energy. The upper curves
are the total cross section for diffractive production while the low
curves show the energy behavior of the low mass contribution to the
diffractive cross section from \protect\eq{SDXSSM}. Notations are
the same as in \protect\fig{xsel}.} \label{sdxs}}

However, one can see from \fig{sdxs}  that our approach cannot
reproduce  $\sigma_{SD}$ in the region of low energies (small values
of produced mass).   There are at least two reasons for such a
failure: first, we do not take into account the exchange of the
secondary Reggeons whose contribution is essential,  as we see on the
example of the energy behavior of the total cross section, and,
second,  the contribution of the one extra gluon emission has been
calculated in leading log(1/x) order, which cannot describe the low
mass diffraction. The experimental data on single diffraction (see
Refs. \cite{CDFDD,GOMO}) can be described assuming,   in addition to
the triple Pomeron vertex, that we modeled by extra gluon emission,
a significant contribution from the Reggeon-Pomeron-Reggeon vertex in
Regge phenomenology,  which we cannot include in our approach.

\subsection{Predictions for the LHC range of energies}

As we have seen, the model is able to describe the available
experimental data quite satisfactorily (and Model B even quite well)
both for soft and hard interactions. Therefore, we can rely on the
model B for the predictions in the LHC energy range. From
\fig{txs}-\fig{sdxs} we see that at the LHC energy we have
$\sigma_{tot} = 101.3\,mb$, $\sigma_{el}=28.84\,mb$ ,  $B_{el} =
18.4 \,GeV^{-2}$ and $\sigma_{diff}=10.5\,mb$ .

\TABLE[ht]{
\begin{tabular}{|l|l|l|l|l|l|l|}
\hline
          &               &                &         &        &  &      \\
Model &  $\sigma_{tot} $
&  $  \sigma_{el} $  & $ \sigma_{diff} $ & $B_{el}$ & $A_{el}(b=0) $ & $A_{el}(b=0) $  \\
  & mb & mb & mb & $ GeV^{-2}$ &(Tevatron) & (LHC)\\ \hline
Our A(3D)  & 83.0 & 23.54   & 10 & 16.67 & 0.94 & 0.98 \\
Our B(2D)  & 101.3 & 28.84   & 10.5 & 18.4 & 0.94 & 0.98 \\
GLM1\cite{SPLAST} & 110.5 & 25.3 & 11.6 & 20.5 & 0.6 & 0.7 \\
GLM2\cite{GLMNEW}& 91.7 & 20.9& 11.8& 17.3 & 0.94 & 0.95\\
 RMK\cite{RMK} & 88.0 (86.3)    & 20.1 (18.1)   & 13.3 (16.1)   & 19  & 0.89 & 0.92\\
\hline
\end{tabular}
\caption{Cross sections and elastic slope at the LHC energy in
different models. For the RMK model, we put two parameterizations and the
value for $B_{el}$ directly from the curve in Fig.18 of
Ref.\cite{RMK}}}

Table 3 shows that our prediction for LHC energy is close to ones
that are given by the models that fitted the experimental data. The
RMK  and GLM2 models are  based on soft Pomeron phenomenology while
the GLM1 model is close to our approach: in this model the existence
of soft Pomeron is not assumed and the parametrization was chosen
for from the same ideas as our approach here. However, in the GLM1
model all parameters are found from fitting soft data. In spite the
fact that the values of cross section are close the different models
are  different in more detailed characteristics. For example, one
can see from the Table 1  they predict different values for
$A_{el}(b =0)$ . From unitarity $ A_{el}(b =0) \,\leq \,1$. Our
model predicts that at the LHC energy our proton-proton collision is
close to the black disc regime. The same we can see in GLM2 model,
but in the RMK model we are not so close to this regime at the LHC,
while in the GLM1 models the proton-proton interaction is far away
from the black disc limit.

As you can see from \eq{txs}-c we have some information on the total
cross section from the cosmic ray experiment. However, to extract
the value of the total cross section from such an experiment, we need to
know the meson-proton cross section as well. In \fig{mespro}, we
plotted our predictions which we hope will be useful for discussing the
cosmic ray experiment\footnote{ We thank S. Nussinov for drawing our
attention to the necessary knowledge of the meson-proton total cross sections
at high energy.}.

\section{Survival probability for diffractive Higgs production}

\FIGURE[h]{ \centerline{\epsfig{file=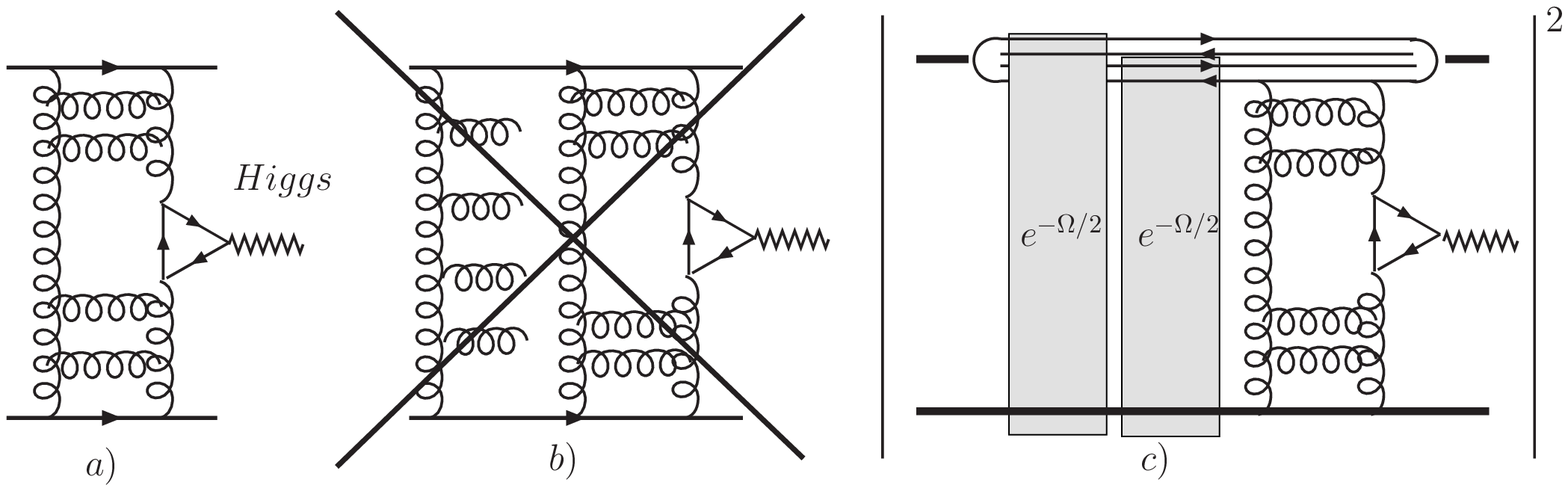,width=140mm}}
\caption{ Diffractive Higgs production.} \label{spH} } The
diffractive Higgs production, is the reaction which has the best
experimental signature for the discovery of the Higgs boson at the LHC.  At
fist sight this process   occurs at short distances of the order of
$1/M_H$, where $M_H$ is the  mass of Higgs boson ($ M_h \geq 100
GeV$) and can be calculated in perturbative QCD (see \fig{spH}-a and
Ref.\cite{DG} for calculations).  However, as was noticed a long ago
\cite{Bj,GLM1} that it is not enough to calculate the diagram of
\fig{spH}-a  which describes the Higgs boson production from one
parton shower. We have  to  multiply this cross section by a
probability that two parton showers (or more) will not interact with
the target since they could produced hadrons that will fill up the large
rapidity gaps between Higgs and protons in the final state (see
\fig{spH}-b).  This  probability we call the survival probability. In
the case of  our model with the eikonal type formula for the scattering
amplitude, the expression for the survival probability is  very
simple (see \fig{spH}-c), namely,\cite{Bj,GLM1} \beq \label{spH1}
\langle|S^2 |\rangle\,\,\,=\,\,\frac{\int d^2 b\,\,\Lb \int d^2 r_1
d^2 r_2 |\Psi_{proton}(r_1,r_2)|^2\,
 e^{ - \h \Omega(s,r_1,b)} \Rb^4\,\sigma_H(b; \fig{spH}-a )}{ \int d^2 b
 \,\sigma_H(b; \fig{spH}-a )}
\eeq
The appearance of the factor $\exp\Lb -\h \Omega(s,r_1,b)\Rb$ in \eq{spH1} is clear from  \fig{spH}-c) but the power of 4 requires discussion. Actually this power reflects the fact that we have to find the probability that neither dipole 1 nor dipole 2 could scatter inelastically. The probability that one dipole does not scatter inelastically is equal to
$$
\Lb \int d^2 r_1 d^2 r_2 |\Psi_{proton}(r_1,r_2)|^2\,
 e^{ - \h \Omega(s,r_1,b)} \Rb^2\,\,\equiv \,\,\Lb \langle| e^{ - \h \Omega(s,r_1,b)} |\rangle \Rb^2
 $$
Therefore for the probability that two dipoles  cannot scatter
inelastically we obtain $\Lb \langle| e^{ - \h \Omega(s,r_1,b)}
|\rangle \Rb^4$. \FIGURE[h]{
\centerline{\epsfig{file=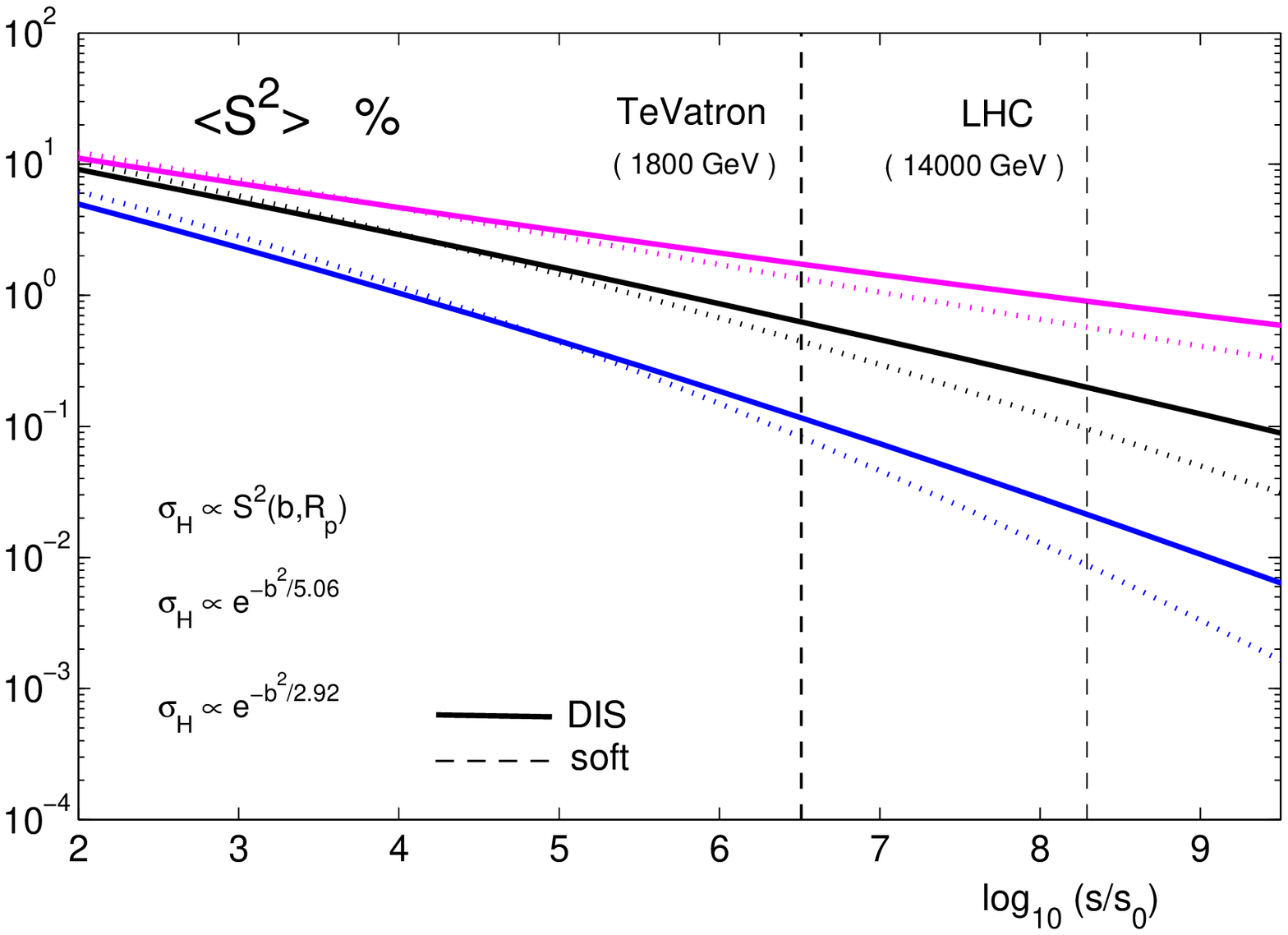,width=120mm,height=90mm}}
\caption{The value of the survival probability for diffractive Higgs
production versus energy. Solid lines show the survival probability
in Model A that describes the  DIS data with a very good $\chi^2/d.o.f  <
1$, while dashed lines correspond to $ <S^2>$ in Model B which gives
a good description of all soft data.} \label{sp} } We can see from
\eq{spH1} that for the calculation of the survival probability we need
to know the $b$ - dependence of the hard cross section. This
dependence has been discussed in detail in Ref.\cite{GKLMP}, where
it was extracted from the process of diffraction production of
J$/\Psi$ with the proton  (elastic) and the state with mass larger than the
proton (inelastic). These two processes have two different slopes in the
$t$ behavior: $B_{el}= 4\, GeV^{-2}$ and $B_{in}= 1. 86
\,GeV^{-2}$\cite{PSISL} . The dependence of the vertex for photon to
J/$\Psi$ transition was extracted from these reactions in Ref.
\cite{KOTE} leading to the values of slopes for the vertices of the
transition of proton to proton (elastic), and proton to non-proton final
state (inelastic): $B_{el}= 3.6\, GeV^{-2}$ and $B_{in}= 1.46 \,
GeV^{-2}$. In our model, the projectile proton first goes to the
state of two free dipoles (see \fig{spH}-c) , these two dipoles can
produce two and more parton showers; and finally two free dipoles
in hard processes create a proton. Therefore, at least the upper vertex is
the same as in the inelastic diffractive production of  J$/\Psi$ in DIS.
Considering the lower vertex being elastic, we can write $\sigma_H(b;
\fig{spH}-a) $ in the form \beq \label{spH2} \sigma_H(b;
\fig{spH}-a)\,\,\,=\,\,\,\frac{\sigma_0}{ \pi R_H}\,e^{-
\frac{b^2}{R^2_H}} \eeq with $R^2_H = B_{in} + B_{el}=5.06\,
GeV^{-2}$. However, it is our model  disadvantage that we describe
differently projectile and target protons. In any case the eikonal
type model that we use,  does not mean that only the elastic
rescattering contributes to the shadowing corrections. Treating both
protons on the same ground we take $R^2_H = 2 B_{in} = 2.92 \,
GeV^{-2}$.

Since our model  naturally includes both soft and hard interactions
we can hope that the hard process of diffractive Higgs production
can be described in the same way as we did in the model using the
same amplitude $\Omega$. However, we need to take into account that
in this case \beq \label{spH3} \sigma_H(b;
\fig{spH}-a)\,\,\,=\,\,\sigma_0 \,S^2(b) \eeq where $S(b)$ is given
by \eq{SB}.

Our results on the survival probability is shown in \fig{sp}, and  in
Table 2. Table 2 demonstrates that our value for the survival
probability turns out to be much  smaller than the values in the
phenomenological models, that fitted the experimental data using the
soft Pomeron approach (see  Table 2, the review of \cite{heralhc} and Ref.
\cite{RMK}). We think that the difference stems not from the details of the
models, but from the key  ingredient of our model: the fact that we
took into account the interaction at short distances.  Therefore, we
confirm that the short distances give a substantial contribution to
the value of survival  probability, as has been noticed in Refs.
\cite{BBK,JM}. The Table 2 also shows that  the value of the
survival probability crucially depends on the  model for the
$b$-dependence of the hard cross section. Our model with \eq{spH3}
for the hard cross section is similar to the one that was used in the GLM
model \cite{SPLAST}. It is interesting that our model confirms the
general tendency to obtain a smaller value for the survival
probability advocated in Ref. \cite{SPLAST}.

\TABLE[ht]{
\begin{tabular}{|c|c|c|}
\hline
          &               &              \\
Model &  $\sigma_{H}(b) $
&  $ \langle|S^2 |\rangle $ \\
 & & \\ \hline
Our Model A(3D)   & \eq{spH2} $R^2_H = 5.06\,GeV^{-2}$ & 0.24\% \\
      & \eq{spH2} $R^2_H = 2.92\,GeV^{-2}$ & 0.02\% \\
   & \eq{spH3} & 0.89\% \\
Our Model B(2D)   & \eq{spH2} $R^2_H = 5.06\,GeV^{-2}$ & 0.24\% \\
      & \eq{spH2} $R^2_H = 2.92\,GeV^{-2}$ & 0.096\% \\
   & \eq{spH3} & 0.57\% \\
GLM1\cite{SPLAST} & Two channel model for $\sigma_H$  & 2\% (0.7\%) \\
GLM2\cite{GLMNEW} & Two channel model  for $\sigma_H$ & 0.21\% \\
 RMK\cite{RMK} &  \eq{spH2} $R^2_H = 11 \,GeV^{-2}$ &  3.2\% (2.3\%)\\
  & \eq{spH2} $R^2_H \,=\, 8 \,GeV^{-2}$ & 1.7\% (1.2\%) \\
\hline
\end{tabular}
\caption{ The survival probability for diffractive Higgs production at
the LHC energy, in different models. In the GLM model, for
$\sigma_H(b)$ is used in the two channel model, with the same $B_{in}$
and $B_{el}$ for the diffractive production of J/$\Psi$ in DIS.}}

\section{Lessons from the model }
Long distance physics is very complicated, non-perturbative
phenomenon, and we certainly do not pretend that we are able to
describe it in its full richness. However, we demonstrate in this
paper that the gap between  this physics and the short distance
physics which  is under full control of perturbative QCD, is not so
huge that it would  be a hopeless task to build a bridge.
Comparing this with the experimental data, we showed that the soft data
depends on the transition region between saturation domain and
perturbative QCD region and, therefore, can give valuable
information on this transition, checking our theoretical approaches
to  it. The widely used, phenomenological soft Pomeron do not appear
in our approach, and we hope that the reader will ask the question: do
we need a soft Pomeron, having in mind our negative answer.

Our description of the experimental data, is not worse (in the case
of Model B)  than the one in the models which fitted the data on the
basis of the soft Pomeron phenomenology (see Refs.
\cite{1CH,2CH,RMK,SPLAST}). Recalling that we fitted all parameters
of our model from DIS processes, we  interpret this success  in the
way that the high energy soft scattering processes are determined by
QCD, at short distances of the order of  $1/Q_s$, where $Q_s$ is the
saturation momentum. Model A describes the data worse than Model B,
but we consider this description quite satisfactory remembering the
crude character of this approach. It should be stressed, once more
that the difference between our model A and model B, is in our
attitude to the simple formula of \eq{MODN}: in model A we trust
\eq{MODN} in the entire kinematic region of accessible distances,
while in model B we view this formula as a kind of qualitative
description, that includes the main features of the saturation
regime. Therefore, we were searching for the parameters of model B, in
the way that describes all the data both on DIS, and on soft interactions,  in
the best possible way.

In our model, we used several assumptions which have a different
theoretical status. The first assumption is the exponential form of
the dipole scattering amplitude (see \eq{EIKN} and \eq{MODN}.   We
have discussed the theoretical arguments for such an assumption,
namely, this form has been proven for the transition region between
perturbative QCD, and the saturation domain (see also Ref. \cite{LMP}).
It should be stressed that the soft data, are sensitive to the
transition region as we have discussed.

The second assumption, is the expression of \eq{OMEGA} for $\Omega\Lb
x,r,b \Rb$. This assumption is a compromise between what we should
do, and what we can do. We can check it by describing the DIS
experimental data, at large values of the photon virtualities, where
the difference between our  formula for $\Omega$, and the DGLAP
expression for it should be large.

The third  assumption is the impact parameter dependence of
$\Omega\Lb x,r,b \Rb$ (see \eq{LRD} and  \eq{SB}). This is a pure
phenomenological ansatz, since the current stage of our theory  does
not allow us to find $b$ dependence \cite{KOWI}. Actually, only soft
data allows us to check this dependence. Indeed, our $\Omega$
describes the $t$-dependence of the elastic cross section while ,
for example,  the Gaussian $b$ dependence results in the appearance of
the structure of maxima and minima, at small values of $t$, which
contradicts the experimental behavior of $d \sigma_{el}/d t$.

In \fig{nle}-a we plot the dipole amplitude averaged with the parton
wave function \beq \label{AVN} \langle| N | \rangle| \,\,\,
\equiv\,\,\, \int d^2 r_1 d^2 r_2 |\Psi_{proton}(r_1,r_2)|^2
N(x,r_1, \vec{b} =0) \eeq

One can see, that this amplitude increases and it approaches 1.
However, it happens at ultra high energies and at the Tevatron
energy, for example, this amplitude is only $0.8$ at $b =0$. Such an
average scattering amplitude, leads to the elastic amplitude of
proton-proton scattering $ 2N - N^2 = 0.96$.  The averaged  slowly
increases with energy, and reaches 1 at energy.   At the LHC we
expect $\langle|N|\rangle =0.915$ and the average  elastic amplitude
is close to unity $\langle|2N - N^2|\rangle = 0.99$. Such a behavior
shows, that the values as well as the dependence on energy crucially
depends on the behavior of our dipole amplitude, in the vicinity of the
saturation scale. To illustrate the strength of the saturation
effect, we plot the average $\Omega$ at $b=$ as a function of energy (see
\fig{nle}-b), which is a considerably overshoots the average scattering
amplitude. It should be stressed, that in spite of the fact that the
elastic amplitude is very close to unity at the LHC energy, one can
see from \fig{nle}-a that   $\langle|N|\rangle$ is only 0.9.  and
the asymptotic behavior with  $\langle|N|\rangle$  close to 1, starts
from $s = 10^{12}\, GeV^2$.

\FIGURE[h]{
\begin{tabular}{c c}
\epsfig{file=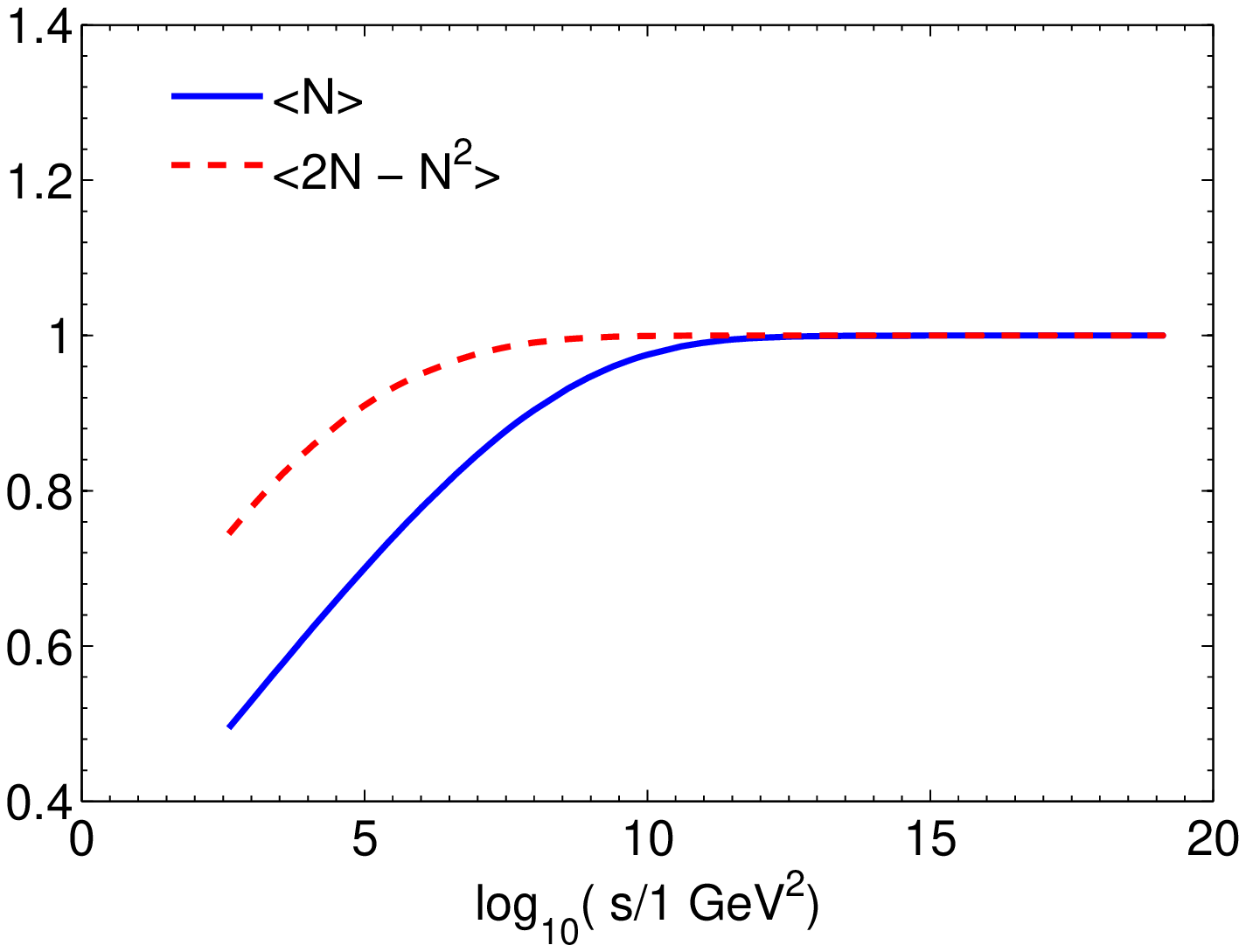,width=75mm,height=60mm} &\epsfig{file=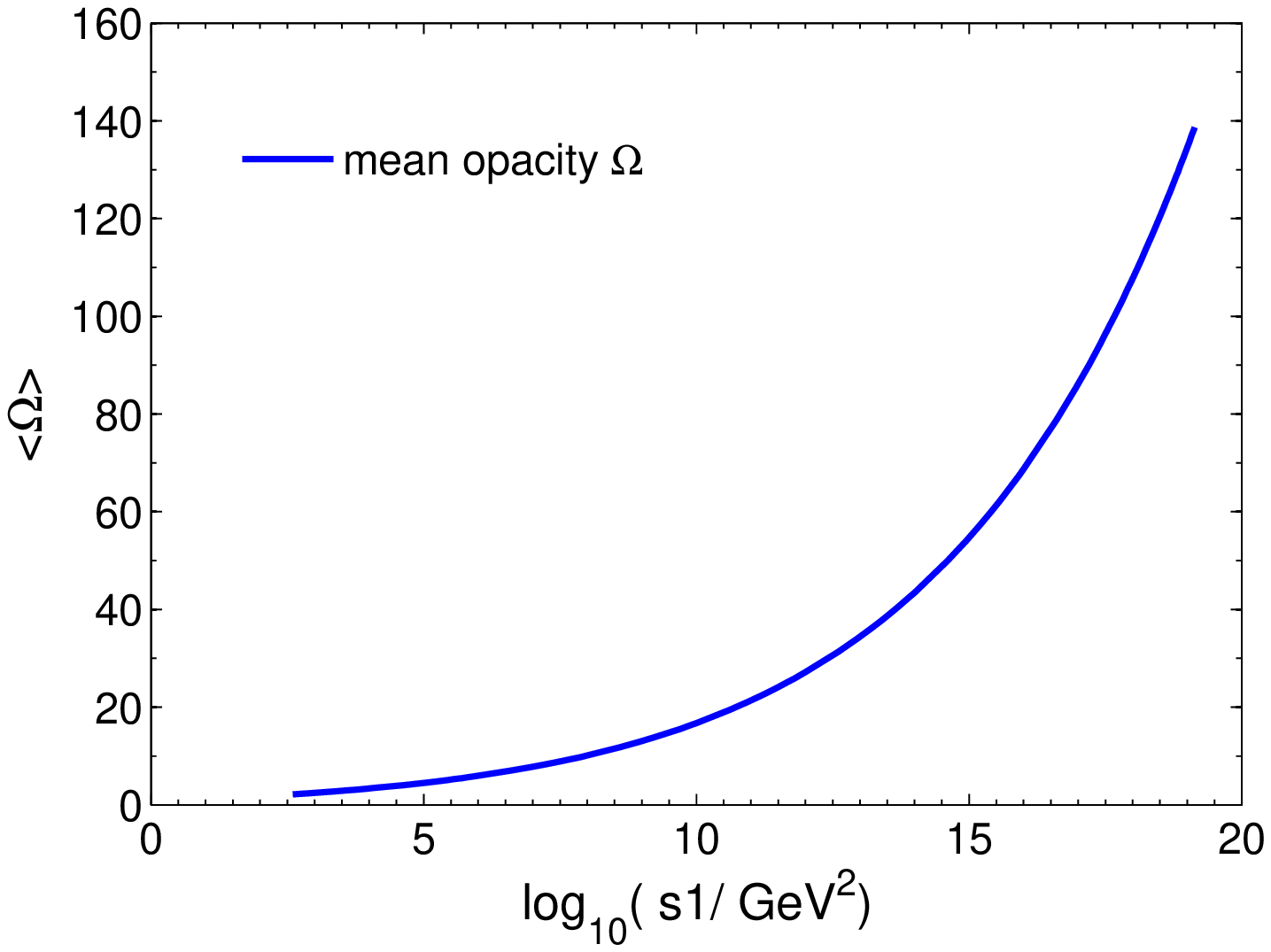,width=75mm,height=60mm}\\
\fig{nle}-a & \fig{nle}-b\\
\end{tabular}
\caption{ Average  dipole amplitude ($\langle{N}\rangle $ see \eq{AVN}), average  proton- proton amplitude (both in \fig{nle}-a) and average opacity $\Omega$(see \fig{nle}-b)  at $b=0$ as a function of energy in our model.}
\label{nle} }

The principle difference between our models and the models of soft
interactions, is the fact that we predict all observables fitting all
parameters from the DIS data.   The GLM1 model\cite{SPLAST} is close
to our approach ideologically, because it does not assume the
existence of the soft Pomeron.  However, at first sight this model has
two major shortcomings: minima at small $t$ which have not been seen
experimentally; and the slow fall down of the cross section of
diffractive production, as a function of the mass of produced system of
hadrons. Our model shows, that the $t$ dependence can be easily heeled by
assuming the exponential form for the profile function, instead of the
Gaussian one that has been used in the GLM1 model. As far as large
mass diffraction that has been neglected in the GLM1
model\cite{SPLAST}, we show that this diffraction is essential. In
this respect, we are close to the RMK model\cite{RMK}  and to the GLM2
model \cite{GLMNEW} which  are  based on the soft Pomeron exchange.
Our model can be considered as an argument that the multi Pomeron
exchanges, and the Pomeron interactions could be essential for high
mass diffraction.

In general  we demonstrated in this paper that the distances,
essential in so called soft interactions, is not so long, but rather about
$1/Q_s \ll R_h$, where $R_h$ is the hadron radius. This conclusion
stems both from the success of our model in the description of the data,
using the parameters fitted in DIS,  and from the use of the energy
variable $x_{soft}$, which is determined by the saturation scale.

We hope that our model will generate deeper theoretical ideas on the
matching between soft and hard interactions, based on high parton
density QCD.

\section* {Acknowledgements}
We are grateful to Jochen Bartels, Errol Gotsman, Lev Lipatov, Uri
Maor and Misha Ryskin  for fruitful discussions on the subject. This
research was supported in part by the Israel Science Foundation,
founded by the Israeli Academy of Science and Humanities, by BSF
grant $\#$ 20004019 and by a grant from Israel Ministry of Science,
Culture and Sport and the Foundation for Basic Research of the
Russian Federation.


\begin{thebibliography}{99}
\bibitem{GLR}
L. V. Gribov, E. M. Levin and M. G. Ryskin, {\it Phys. Rep.}\,
{\bf 100}, 1 (1983).

\bibitem{MUQI}
A. H. Mueller and J. Qiu,  {\it Nucl. Phys.},427 {\bf B 268}
(1986) .
\bibitem{MV}
L. McLerran and R. Venugopalan, {\it  Phys. Rev.}  {\bf D 49},2233,
3352  (1994); {\bf D 50},2225 (1994); {\bf D 53},458 (1996); {\bf
D 59},09400
(1999).

\bibitem{DGLAP}
 V. N. Gribov and L. N. Lipatov, {\it  Sov. J. Nucl. Phys}  {\bf 15} (1972)
                438;\\
 G. Altarelli and G. Parisi,{\it  Nucl. Phys.}\, {\bf B 126} (1977) 298; \\
Yu. l. Dokshitser, {\it Sov. Phys. JETP}  {\bf 46}  (1977) 641.


\bibitem{BFKL}
 E. A. Kuraev, L. N. Lipatov, and F. S. Fadin, {\it  Sov. Phys.
JETP}
                {\bf 45}, 199 (1977); \,\,\,
Ya. Ya. Balitsky and L. N. Lipatov,
               {\it   Sov. J. Nucl. Phys.}\, {\bf 28}, 22 (1978).

\bibitem{BK}
I.~Balitsky,
[arXiv:hep-ph/9509348];\,\,
{\it Phys.\ Rev.} {\bf D60}, 014020 (1999)
[arXiv:hep-ph/9812311]\,\,\,\,
Y.~V.~Kovchegov,
{\it Phys.\ Rev.}  {\bf D60}, 034008  (1999),
[arXiv:hep-ph/9901281].
\bibitem{JIMWLK}
~J.~Jalilian-Marian, A.~Kovner, A.~Leonidov and H.~Weigert,
{\it  Phys.\ Rev.}\,  {\bf D59}, 014014 (1999),
[arXiv:hep-ph/9706377];\,\,  {\it Nucl.\ Phys.}\,{\bf B504}, 415
(1997),
[arXiv:hep-ph/9701284]; \,\,\,
J.~Jalilian-Marian, A.~Kovner and H.~Weigert,
  {\it Phys.\ Rev.}  {\bf D59}, 014015 (1999),
  [arXiv:hep-ph/9709432];\,\,\,
 A.~Kovner, J.~G.~Milhano and H.~Weigert,
 {\it  Phys.\ Rev.}  {\bf D62}, 114005 (2000),
  [arXiv:hep-ph/0004014]\,; \,\,\,
E.~Iancu, A.~Leonidov and L.~D.~McLerran,
{\it  Phys.\ Lett.}\,  {\bf B510}, 133 (2001);
[arXiv:hep-ph/0102009];\,\, {\it  Nucl.\ Phys.}\,  {\bf A692}, 583
(2001),
[arXiv:hep-ph/0011241];\,\,\,
E.~Ferreiro, E.~Iancu, A.~Leonidov and L.~McLerran,
 {\it  Nucl.\ Phys.}\  {\bf A703}, 489 (2002),
  [arXiv:hep-ph/0109115];\,\,\,
H.~Weigert,
{\it  Nucl.\ Phys.}  {\bf A703}, 823 (2002),
[arXiv:hep-ph/0004044].

\bibitem{KOLU}
  A.~Kovner and M.~Lublinsky,
 {\it  Phys.\ Rev.}\, {\bf D  71}, 085004 (2005)
  [arXiv:hep-ph/0501198].

\bibitem{HIMST}
  Y.~Hatta, E.~Iancu, L.~McLerran, A.~Stasto and D.~N.~Triantafyllopoulos,
 {\it  Nucl.\ Phys.}\  {\bf A764}, 423 (2006)
  [arXiv:hep-ph/0504182].

\bibitem{STPH}
 E.~Iancu, A.~H.~Mueller and S.~Munier,
{\it   Phys.\ Lett.} \, {\bf B606} (2005) 342
  [arXiv:hep-ph/0410018];
E.~Brunet, B.~Derrida, A.~H.~Mueller and S.~Munier,
  arXiv:cond-mat/0603160;
{\it   Phys.\ Rev.}\,  {\bf E73} (2006) 056126
  [arXiv:cond-mat/0512021].

\bibitem{EGM}
 R.~Enberg, K.~Golec-Biernat and S.~Munier,
  {\it Phys.\ Rev.}\  {\bf D72} (2005) 074021
  [arXiv:hep-ph/0505101].
S.~Munier,
 {\it  Phys.\ Rev.}\ ,  {\bf D 75} (2007) 034009
  [arXiv:hep-ph/0608036].

\bibitem{LMP}
 E.~Levin, J.~Miller and A.~Prygarin,
  {\it ``Summing Pomeron loops in the dipole approach,''} Nucl.Phys. A ({\it in press});
  arXiv:0706.2944 [hep-ph].

\bibitem{NS}
N.~Armesto and M.~A.~Braun,
  {\it Eur.\ Phys.\ J.}\  {\bf C20}, 517 (2001)
  [arXiv:hep-ph/0104038];\,\,
M.~Lublinsky,
  {\it Eur.\ Phys.\ J.}\  {\bf C21}, 513 (2001)
  [arXiv:hep-ph/0106112];\,\,\,\,
E.~Levin and M.~Lublinsky,
 {\it   Nucl.\ Phys.}  {\bf A712}, 95 (2002)
  [arXiv:hep-ph/0207374];\,\,
  {\it Nucl.\ Phys.}  {\bf A712}, 95 (2002)
  [arXiv:hep-ph/0207374];\,\,
{\it Eur.\ Phys.\ J.}\, {\bf C22}, 647 (2002)
  [arXiv:hep-ph/0108239];\,\,\,\,
M.~Lublinsky, E.~Gotsman, E.~Levin and U.~Maor,
 {\it   Nucl.\ Phys.}\,  {\bf A696}, 851 (2001)
  [arXiv:hep-ph/0102321];\,\,
 {\it  Eur.\ Phys.\ J.}\,  {\bf C27}, 411 (2003)
  [arXiv:hep-ph/0209074];\,\,\,\,
K.~Golec-Biernat, L.~Motyka and A.Stasto,
 {\it  Phys.\ Rev.} \, {\bf D65}, 074037 (2002)
  [arXiv:hep-ph/0110325];\,\,\,
E.~Iancu, K.~Itakura and S.~Munier, {\it
  Phys.\ Lett.}\, {\bf B590} (2004) 199
  [arXiv:hep-ph/0310338].
K.~Rummukainen and H.~Weigert,
{\it   Nucl.\ Phys.} \, {\bf A739}, 183 (2004)
  [arXiv:hep-ph/0309306];\,\,K.~Golec-Biernat and A.~M.~Stasto,
 {\it  Nucl.\ Phys.} {\bf B668}, 345 (2003)
  [arXiv:hep-ph/0306279];\,\,\,\,E.~Gotsman, M.~Kozlov, E.~Levin, U.~Maor and E.~Naftali,
 {\it   Nucl.\ Phys.}\, {\bf A742}, 55 (2004)
  [arXiv:hep-ph/0401021];\,\,\,\,K.~Kutak and A.~M.~Stasto,
 {\it  Eur.\ Phys.\ J.}\,  {\bf C41}, 343 (2005)
  [arXiv:hep-ph/0408117];\,\,\,\,G.~Chachamis, M.~Lublinsky and A.~Sabio Vera,
{\it   Nucl.\ Phys.}  {\bf A748}, 649 (2005)
  [arXiv:hep-ph/0408333];\,\,\,\,
 J.~L.~Albacete, N.~Armesto, J.~G.~Milhano, C.~A.~Salgado and U.~A.~Wiedemann,
 {\it  Phys.\ Rev.} {\bf D71}, 014003 (2005)
  [arXiv:hep-ph/0408216];\,\,\,\,E.~Gotsman, E.~Levin, U.~Maor and E.~Naftali,
{\it   Nucl.\ Phys.} \,{\bf A750} (2005) 391
  [arXiv:hep-ph/0411242].


\bibitem{MOD}
   K.~J.~Golec-Biernat and M.~Wusthoff,
 {\it  Phys.\ Rev.}\,   {\bf D  59} (1999) 014017
  [arXiv:hep-ph/9807513]\,\,\,\.,
 {\it  Phys.\ Rev.}\,   {\bf D  60} (1999) 114023
  [arXiv:hep-ph/9903358]\,;\,\,\,
    E.~Gotsman, E.~Levin, M.~Lublinsky, U.~Maor, E.~Naftali and K.~Tuchin,
  {\it J.\ Phys.}\, {\bf G 27} (2001) 2297
  [arXiv:hep-ph/0010198]\,;\,\,\,
    H.~Kowalski and D.~Teaney,
 {\it  Phys.\ Rev.}\,   {\bf D  68} (2003) 114005
  [arXiv:hep-ph/0304189]\,;\,\,
    J.~Bartels, K.~J.~Golec-Biernat and H.~Kowalski,
 {\it  Phys.\ Rev.}\,   {\bf D 66} (2002) 014001
  [arXiv:hep-ph/0203258].

\bibitem{KOR}
 A.~Kormilitzin,
  {\it ``Saturation model in the non-Glauber approach,''}
  arXiv:0707.2202 [hep-ph].

\bibitem{Adloff:2000qk}
  C.~Adloff {\it et al.}  [H1 Collaboration],
  Eur.\ Phys.\ J.\  C {\bf 21}, 33 (2001)
  [arXiv:hep-ex/0012053].

\bibitem{Chekanov:2001qu}
  S.~Chekanov {\it et al.}  [ZEUS Collaboration],
  Eur.\ Phys.\ J.\  C {\bf 21}, 443 (2001)
  [arXiv:hep-ex/0105090].
\bibitem{Breitweg:2000yn}
  J.~Breitweg {\it et al.}  [ZEUS Collaboration],
  Phys.\ Lett.\  B {\bf 487}, 53 (2000)
  [arXiv:hep-ex/0005018].

\bibitem{BALE}
 J.~Bartels and E.~Levin,
 {\it  Nucl.\ Phys.}\,   {\bf B387} (1992) 617.
\bibitem{GS}
  ~J.~Kwiecinski and A.~M.~Stasto,
{\it  Acta Phys.\ Polon.}\, {\bf B33} (2002) 3439;\,\,{\it Phys.\ Rev.}\,
{\bf
D66}
(2002) 014013
[arXiv:hep-ph/0203030];\,\,\,\,\,A.~M.~Stasto, K.~Golec-Biernat and
J.~Kwiecinski,
{\it  Phys.\ Rev.\ Lett.}\,  {\bf 86} (2001) 596
arXiv:hep-ph/0007192].

\bibitem{IIM}
E.~Iancu, K.~Itakura and L.~McLerran,
 {\it  Nucl.\ Phys.}\,    {\bf A708} (2002) 327
  [arXiv:hep-ph/0203137].


\bibitem{REV}
L.~McLerran,
  {\it ``Some comments about the high energy limit of QCD,''}
  { \it Acta Phys.\ Polon.}\,  {\bf B 37} (2006) 3237
  [arXiv:hep-ph/0702] and references therein;\,\,\,
  E.~Ferreiro, E.~Iancu, K.~Itakura and L.~McLerran,
  { \it Nucl.\ Phys.}\,   {\bf A 710} (2002) 373
  [arXiv:hep-ph/0206241];\,\,\,
   E.~Iancu, A.~Leonidov and L.~McLerran,
  {\it ``The colour glass condensate: An introduction,''}
  arXiv:hep-ph/0202270;\,\,\,
   E.~Levin,
  {\it ``Saturation 2005 (mini-review),''}
  AIP Conf.\ Proc.\  {\bf 792} (2005) 536
  [arXiv:hep-ph/0506161];
  {\it ``An introduction to pomerons,''}
  arXiv:hep-ph/9808486;\,\,
  E.~Laenen and E.~Levin,
  {\it ``Parton Densities At High-Energy,''}
{\it   Ann.\ Rev.\ Nucl.\ Part.\ Sci.}\,  {\bf 44} (1994) 199;\,\,\,
  E.~M.~Levin and M.~G.~Ryskin,
  {\it ``High-Energy Hadron Collisions In QCD,''}
 {\it  Phys.\ Rept.}\,  {\bf 189} (1990) 267.

\bibitem{DFK}
H.~G.~Dosch, E.~Ferreira and A.~Kramer,
  {\it Phys.\ Rev.}\,   {\bf D  50} (1994) 1992
  [arXiv:hep-ph/9405237] and references therein.

\bibitem{BATAV}
 J.~Bartels, E.~Gotsman, E.~Levin, M.~Lublinsky and U.~Maor,
  {\it Phys.\ Rev.}\,   {\bf D 68} (2003) 054008
  [arXiv:hep-ph/0304166]; 
  {\it Phys.\ Lett.}\,  {\bf B 556} (2003) 114
  [arXiv:hep-ph/0212284].
\bibitem{SPLAST}
E.~Gotsman, E.~Levin and U.~Maor,
  {\it ``A Soft Interaction Model at Ultra High Energies: Amplitudes, Cross Sections
  and Survival Probabilities,''}
  arXiv:0708.1506 [hep-ph].

\bibitem{QS}
S.~Munier and R.~Peschanski,
{\it ``Universality and tree structure of high energy QCD,''}
arXiv:hep-ph/0401215;\,\,{\it Phys.\ Rev.}\,  {\bf D69} (2004) 034008
[arXiv:hep-ph/0310357];\,\,
{\it Phys.\ Rev.\ Lett.}\,  {\bf 91} (2003) 232001
[arXiv:hep-ph/0309177];\,\,\,
A. H. Mueller and V.~N.~Triantafyllopoulos,
{\it Nucl.Phys.} {\bf B640}, 331 (2002);\,\,\,
D.~N.~Triantafyllopoulos,
Nucl.\ Phys.\ B {\bf 648}, 293 (2003).

\bibitem{GESC}
J.~Bartels and E.~Levin,
{\it Nucl.\ Phys.}\,  {\bf B387} (1992) 617;\,\,\,
A. M. Stasto, K. Golec-Biernat and J. Kwiecinski,
{\it Phys. Rev. Lett.}, {\bf 86}, 596 (2001);\,\,\,
~E.~Levin and K.~Tuchin, {\it Nucl.\ Phys.}\, {\bf A693} (2001) 787,
[arXiv:hep-ph/0101275]\,;\,\,{\bf A691}  (2001) 779,[arXiv:hep-ph/0012167];\,
{\bf B573} (2000) 833,  [arXiv:hep-ph/9908317];\,\,\,
E. Iancu, K.~Itakura and L.~McLerran,
{\it Nucl.\ Phys.}  {\bf A708}, 327 (2002).


\bibitem{KOWI}
 A.~Kovner and U.~A.~Wiedemann,
  Phys.\ Lett.\  B {\bf 551} (2003) 311
  [arXiv:hep-ph/0207335]; Phys.\ Rev.\  D {\bf 66} (2002) 034031
  [arXiv:hep-ph/0204277]; Phys.\ Rev.\  D {\bf 66} (2002) 051502
  [arXiv:hep-ph/0112140].

\bibitem{BST}
  R.~C.~Brower, J.~Polchinski, M.~J.~Strassler and C.~I.~Tan,
  JHEP {\bf 0712} (2007) 005
  [arXiv:hep-th/0603115].

\bibitem{LAT}
C.J. Morninstar and M. J. Peardon, Phys. \,Rev. {\bf D60} (2005) 344,
 [arXiv:hep-lat/9901004].
\bibitem{MUCD}
A.~H.~Mueller, {\it Nucl.\ Phys.} {\bf B415}, 373 (1994);
{\it ibid}  {\bf B437}, 107 (1995).


\bibitem{L1}
E.~Levin and M.~Lublinsky,
{\it   Nucl.\ Phys.}\, {\bf A730}, 191 (2004)
  [arXiv:hep-ph/0308279].
\bibitem{L2}
E.~Levin and M.~Lublinsky,
{\it   Phys.\ Lett.} {\bf B607}, 131 (2005)
  [arXiv:hep-ph/0411121].





\bibitem{LT}
E.~Levin and K.~Tuchin,
{\it Nucl.\ Phys.}\  {\bf A693} (2001) 787
[arXiv:hep-ph/0101275];\,\,\,{\bf A691} (2001) 779
[arXiv:hep-ph/0012167];\,\,\,{\bf B573} (2000) 833
[arXiv:hep-ph/9908317].




\bibitem{MSHW}
  A.~H.~Mueller, A.~I.~Shoshi and S.~M.~H.~Wong,
 {\it  Nucl.\ Phys.}\,  {\bf B 715}, 440 (2005)
  [arXiv:hep-ph/0501088].

\bibitem{LELU}
  E.~Levin and M.~Lublinsky,
 {\it  Nucl.\ Phys.}\,  {\bf A 763}, 172 (2005)
  [arXiv:hep-ph/0501173].



\bibitem{IT}
E.~Iancu and D.~N.~Triantafyllopoulos,
{\it   Nucl.\ Phys.}\  {\bf A756}, 419 (2005)
  [arXiv:hep-ph/0411405];\,\,{\it Phys.\ Lett.} {\bf B610}, 253 (2005)
  [arXiv:hep-ph/0501193].






\bibitem{QXS}
E.~Gotsman, E.~Levin and U.~Maor,
 {\it  Nucl.\ Phys.}\,   {\bf B 464} (1996) 251
  [arXiv:hep-ph/9509286] and references therein.
\bibitem{EKL}
K. Ellis, Z. Kunst and E. Levin:
{\it Phys. Rev.}\, {\bf D 50} (1994) 1992.
\bibitem{ER}
G.G. Simon et al.,{\it  Z. Naturforschung}\, {\bf 35 A} (1980)  1;\,\,S.R. Amendola et al.,{\it  Nucl. Phys.}\, {\bf B 277} (1985) 168;\,{\it Phys.Lett.}\, {\bf B 178} (1986) 435.

\bibitem{GOWA}
  M.~L.~Good and W.~D.~Walker,
  Phys.\ Rev.\  {\bf 120} (1960) 1857.
\bibitem{LNCB}
E.~E.~Jenkins,
  Ann.\ Rev.\ Nucl.\ Part.\ Sci.\  {\bf 48} (1998) 81
  [arXiv:hep-ph/9803349] and references therein.
  \bibitem{KOTU}
Y.~V.~Kovchegov and K.~Tuchin,
  Phys.\ Rev.\  D {\bf 65} (2002) 074026
  [arXiv:hep-ph/0111362].
\bibitem{DIF}
C.~Marquet,
  Phys.\ Rev.\  D {\bf 76} (2007) 094017
  [arXiv:0706.2682 [hep-ph]];\,\,\,\,
 H.~Kowalski, L.~Motyka and G.~Watt,
  Phys.\ Rev.\  D {\bf 74}, 074016 (2006)
  [arXiv:hep-ph/0606272];\,\,\,
Y.~V.~Kovchegov,
  Phys.\ Rev.\  D {\bf 64}, 114016 (2001)
  [Erratum-ibid.\  D {\bf 68}, 039901 (2003)]
  [arXiv:hep-ph/0107256];
Y.~V.~Kovchegov and L.~D.~McLerran,
  Phys.\ Rev.\  D {\bf 60}, 054025 (1999)
  [Erratum-ibid.\  D {\bf 62}, 019901 (2000)]
  [arXiv:hep-ph/9903246];\,\,\,
Y.~V.~Kovchegov,
  Phys.\ Rev.\  D {\bf 64}, 114016 (2001)
  [Erratum-ibid.\  D {\bf 68}, 039901 (2003)]
  [arXiv:hep-ph/0107256];\,\,\, J.~R.~Forshaw, R.~Sandapen and G.~Shaw,
  Phys.\ Lett.\  B {\bf 594}, 283 (2004)
  [arXiv:hep-ph/0404192];\,\,\,
  Y.~V.~Kovchegov and L.~D.~McLerran,
  Phys.\ Rev.\  D {\bf 60}, 054025 (1999)
  [Erratum-ibid.\  D {\bf 62}, 019901 (2000)]
  [arXiv:hep-ph/9903246];\,\,\,
  M.~Wusthoff,
  Phys.\ Rev.\  D {\bf 56}, 4311 (1997)
  [arXiv:hep-ph/9702201];\,\,\,
  K.~J.~Golec-Biernat and M.~Wusthoff,
  Phys.\ Rev.\  D {\bf 59}, 014017 (1999)
  [arXiv:hep-ph/9807513].
E.~Levin and M.~Wusthoff,
  Phys.\ Rev.\  D {\bf 50}, 4306 (1994).

\bibitem{DL}
A. Donnachie and P.V. Landshoff,
{\it Nucl. Phys.} {\bf B231}, (1984) 189;
{\it Phys. Lett.} {\bf B296}, (1992) 227;
{\it Zeit. Phys.} {\bf C61}, (1994) 139.

\bibitem{DG}
V.~A.~Khoze, A.~D.~Martin and M.~G.~Ryskin,
  {\it Phys.\ Lett.}\,   {\bf B  650} (2007) 41
  [arXiv:hep-ph/0702213];
 \,\,\,
  {\it Eur.\ Phys.\ J.}\,   {\bf C 24} (2002) 459
  [arXiv:hep-ph/0201301];\,\,\,
{\it   Eur.\ Phys.\ J.}\,  {\bf C 26} (2002) 229
  [arXiv:hep-ph/0207313]\,\,\,
  {\it Eur.\ Phys.\ J.}\,   {\bf  C 14} (2000) 525
  [arXiv:hep-ph/0002072];\,\,
\bibitem{Bj}
J. D. Bjorken,
{\it Int. J. Mod. Phys.} {\bf A7}, (1992) 4189;
{\it Phys. Rev.} {\bf D47}, (1993) 101.
\bibitem{GLM1}
E. Gotsman, E.M. Levin and U. Maor,
{\it Phys. Lett.} {\bf B309}, (1993) 199.
\bibitem{GKLMP}
E. Gotsman, H. Kowalski, E. Levin, U. Maor and A. Prygarin,
{\it Eur. Phys. J.} {\bf C47}, (2006) 655.
\bibitem{PSISL}
ZEUS Collaboration,
{\it Nucl. Phys.} {\bf B695} (2004) 3;
{\it Eur. Phys. J.} {\bf C24} (2002) 345.
\bibitem{KOTE}
    H.~Kowalski and D.~Teaney,
  {\it Phys.\ Rev.}\,   {\bf D 68} (2003) 114005
  [arXiv:hep-ph/0304189].
\bibitem{GOMO}
K. Goulianos and J. Montanha, {\it Phys. Rev.} {\bf D59} (1999) 114017.
\bibitem{CDFDD}
F. Abe et al.(CDF  Collaboration),    {\it Phys. Rev.} {\bf D50} (1994)  5535.

\bibitem{heralhc}
E. Gotsman, E. Levin, U. Maor, E. Naftali and A. Prygarin,
{\it "HERA and the LHC - A workshop on the implications of
HERA for LHC  physics: Proceedings Part A"} (2005) 221.
(arXiv:hep-ph/0511060[hep-ph]).
\bibitem{RMK}
M.~G.~Ryskin, A.~D.~Martin and V.~A.~Khoze,
  {\it ``Soft diffraction at the LHC: a partonic interpretation,''}
  arXiv:0710.2494 [hep-ph].
\bibitem{GLMNEW}
  E.~Gotsman, E.~Levin, U.~Maor and J.~S.~Miller,
  {\it``A QCD motivated model for soft interactions at high energies,''}
  arXiv:0805.2799 [hep-ph].
\bibitem{BBK}
 J. Bartels, S. Bondarenko, K. Kutak and L. Motyka,
{\it Phys. Rev.}  {\bf D73} (2006) 093004.
\bibitem{JM}
J. S. Miller,
{\it ``Survival probability in diffractive Higgs production in high density QCD,''}
{\it Eur. Phys. J.\,(in press)}
  arXiv:hep-ph/0610427.
\bibitem{2CH}
E. Gotsman, E. Levin and U. Maor,
{\it Phys. Lett.} {\bf B452}, (1999) 387.
\bibitem{1CH}
E. Gotsman, E. Levin and U. Maor,
{\it Phys. Rev.} {\bf D49}, (1994) R4321


\bibitem{RY}
I. Gradstein and I. Ryzhik, {\it `` Tables of Series, Products, and
Integrals"}, Verlag MIR, Moskau,1981.

\end{thebibliography}
\end{document}